\documentclass[lettersize,journal]{IEEEtran}
\pdfoutput=1
\usepackage{amsmath,amsfonts}
\usepackage{algorithmic}
\usepackage{algorithm}
\usepackage{array}
\usepackage{textcomp}
\usepackage{stfloats}
\usepackage{url}
\usepackage{verbatim}
\usepackage{graphicx}
\usepackage{float}
\usepackage{cite}
\usepackage{amssymb}
\usepackage{subfigure}
\usepackage{color} 

\begin{document}

\title{Cooperative ISAC-empowered \\Low-Altitude Economy}

%
\author{Jun Tang, Yiming Yu, Cunhua Pan, Hong Ren, Dongming Wang, Jiangzhou Wang,~\IEEEmembership{Fellow,~IEEE} and \\
	Xiaohu You,~\IEEEmembership{Fellow,~IEEE}
	
\thanks{
	\textcolor{black}{Jun Tang, Cunhua Pan, Hong Ren, Dongming Wang, Jiangzhou Wang and Xiaohu You are with National Mobile Communications Research Laboratory, Southeast University, Nanjing, China. (e-mail: 220241016, cpan, hren, wangdm, j.z.wang, xhyu@seu.edu.cn).}
	
	Yiming Yu is with the Department of Radio Engineering, China Mobile Group Design Institute Company Ltd., Beijing, China. (e-mail:yuyiming@cmdi.chinamobile.com).
	}

\thanks{\itshape{Corresponding author:} Cunhua Pan.}
}



\maketitle

\begin{abstract}
This paper proposes a cooperative integrated sensing and communication (ISAC) scheme for the low-altitude sensing scenario, aiming at estimating the parameters of the unmanned aerial vehicles (UAVs) and enhancing the sensing performance via cooperation. The proposed scheme consists of two stages. In Stage I, we formulate the monostatic parameter estimation problem via using a tensor decomposition model. By leveraging the Vandermonde structure of the factor matrix, a spatial smoothing tensor decomposition scheme is introduced to estimate the UAVs' parameters. \textcolor{black}{To further reduce the computational complexity, we design a reduced-dimensional (RD) angle of arrival (AoA) estimation algorithm based on generalized Rayleigh quotient (GRQ).} In Stage II, the positions and true velocities of the UAVs are determined through the data fusion across multiple base stations (BSs). Specifically, we first develop a false removing minimum spanning tree (MST)-based data association method to accurately match the BSs' parameter estimations to the same UAV. Then, a Pareto optimality method and a residual weighting scheme are developed to facilitate the position and velocity estimation, respectively. We further extend our approach to the dual-polarized system. Simulation results validate the effectiveness of the proposed schemes in comparison to the conventional techniques.
\end{abstract}

\begin{IEEEkeywords}
Integrated sensing and communication (ISAC), cooperative sensing, tensor decomposition, parameter estimation, data fusion. 
\end{IEEEkeywords}

\section{Introduction}
In the forthcoming fifth-generation advanced (5G-A) and sixth-generation (6G) mobile communication systems, the plethora of low-altitude applications in transportation, tourism, agriculture and emergency services boosts the burgeoning of low-altitude economy (LAE)\cite{9456851}. In LAE, the interaction between the unmanned aerial vehicles (UAVs) and the base stations (BSs) is critical. On one hand, the UAVs need to communicate with the BSs for data transmission. On the other hand, the BSs are also supposed to sense and detect the UAVs to prevent the unauthorized intrusion or collision. To this end, the novel paradigm known as integrated sensing and communication (ISAC) is recognized as a promising solution. Compared to the conventional radar deployment, ISAC systems can simultaneously achieve the superior sensing capabilities and high data transmission rates\cite{9737357}. 

In the low-altitude sensing scenarios, depending on the spatial deployment of the transmitter and the receiver, ISAC configurations can be classified into three primary categories:
monostatic\cite{9906898, 9724260}, bistatic \cite{8318564,9860521} and multistatic \cite{han2024cellular,10464728}. \textcolor{black}{Specifically, in the bistatic and multistatic systems, one or several BSs acts as the transmitters with others served as the receivers to receive the sensing symbols. However, these configurations impose high requirements on the deployment locations and the synchronization across multiple BSs. Differently, in the monostatic setup, the BS transmits the sensing symbols simultaneously while receiving the echo signals reflected by the UAVs, which simplifies the deployment of BS and enhances the system's tolerance to asynchronization.} However, conventional monostatic configuration still faces several challenges. First, due to the path loss, the sensing performance for the distant UAVs is relatively poor\cite{9916293}. In addition, if there exists the obstructions (such as the tall buildings or trees) between the UAVs and the BSs, the detection probability will decrease dramatically. \textcolor{black}{Moreover, the comprehensive status of the UAVs, such as their true velocities cannot be fully captured, since only the radial velocities can be derived from the echo signals\cite{han2024cellular}.}

To overcome the above limitations of the conventional monostatic configuration, the cooperative ISAC framework was proposed\cite{10032141,10207026,9724258,zhang2024,10226276}. 
In the cooperative ISAC networks, each BS first estimates the targets' parameters from the received echo signals. Then, the estimated parameters are fused in the cloud to enhance the sensing performance. \textcolor{black}{In this scenario, targets that are far from a certain BS may be relatively close to other BSs, thereby enhancing the detection probability and the coverage of the ISAC network. Moreover, more diverse sensing directions and richer targets' information brought by the distributed deployment of BSs further boosts the sensing performance, thereby reducing the estimation error and enabling the ability to recover more comprehensive information of the UAVs, such as their true velocities in the three-dimensional (3-D) space.} Recent studies in the field of cooperative ISAC have aimed at designing the sensing solutions to facilitate the sensing capabilities in various scenarios. \textcolor{black}{In the single-antenna setup, \cite{9724258,zhang2024} designed the two-stage target localization schemes in the orthogonal frequency division multiplexing (OFDM) network, where the ranges of targets were respectively derived via the compressed sensing (CS) and the two-dimensional (2-D) fast Fourier transform (FFT) algorithms, then the targets' positions were estimated based on the range estimations. Likewise, \cite{10226276} first derived the target's range and radial velocity, and then proposed a symbol-level fusion scheme across multiple BSs to facilitate the position and velocity estimation. In the multi-antenna setup, in addition to the ranges and velocities, the BSs are also supposed to capture the angles of arrival (AoAs) of the targets to further enhance the sensing performance, and several classic AoA estimation techniques such as multiple signal classification (MUSIC) algorithm have been adopted to derive the targets' AoAs in ISAC systems\cite{10615952}. Beyond the aforementioned conventional estimation methods, tensor techniques are gaining increasing popularity for extracting the multi-dimensional parameters recently, and have been applied for the channel and target parameter estimation due to its superior estimation accuracy and the guaranteed automatic parameter pairing\cite{7914672,9049103,wang2024,10403776}. The authors of \cite{10403776} delved into a preliminary exploration of the tensor-based ISAC, where the channel and target parameter estimation were achieved through a unified tensor approach. However, \cite{10403776} did not reveal the benefits of improving the sensing accuracy through the multi-BS cooperation.}  

As previously mentioned, although the above studies have preliminarily demonstrated the potential of the cooperative ISAC configuration, \textcolor{black}{most of the existing studies only considered some special scenarios of the cooperative ISAC networks (such as the single-antenna or the single-target cases), lacking the generality to extend to the more complex and practical scenarios. Specifically, for the monostatic parameter estimation, several estimation algorithms only considered the single-target scenarios, when extended to the multi-target scenario, how to match the multi-dimensional parameters to the same target still needs to be well designed. For the multi-BS cooperation, the data association across BSs is another critical issue in the multi-target scenario. Although several association methods have been developed \cite{9724258, zhang2024}, they are merely designed for the single-antenna setup. To address these issues, this paper aims to provide a comprehensive scheme for a more general multi-antenna multi-target sensing scenario within the cooperative ISAC framework.} The main contributions of this work are summarized as follows: 
\begin{enumerate}
	\item \textcolor{black}{We consider a general cooperative ISAC system, aiming at sensing multiple UAVs via the cooperation across the multiple multi-antenna BSs. To address the challenging sensing issue, a comprehensive scheme is developed.}
	
	\item \textcolor{black}{First, several preliminary steps are presented to provide some prior and guidelines for the subsequent sensing scheme.} Then, the monostatic parameter estimation problem is formulated via using a tensor decomposition model. By leveraging the Vandermonde structure of the factor matrix, an spatial smoothing tensor decomposition algorithm is developed to derive the AoAs, ranges, radial velocities, and channel coefficients of the UAVs. \textcolor{black}{Additionally, we develop a reduced-dimensional (RD) AoA estimation algorithm based on generalized Rayleigh quotient (GRQ) to further reduce the complexity. }
	
	\item Subsequently, a false removing minimum spanning tree (MST)-based multi-BS data association method is presented.  Compared to the conventional exhaustive permutation method, this approach not only prevents the false detections from impacting the subsequent data fusion, but also reduces the complexity effectively when there are a large number of UAVs.
	
	\item Finally, a Pareto optimality method and a residual weighting scheme are presented to facilitate the
	position and velocity estimation, respectively.
	
	\item We also extend our approach to the dual-polarized system \textcolor{black}{via a forth-order tensor decomposition formulation}, which improves the estimation accuracy when the array size is limited.
\end{enumerate}

The remainder of this paper is organized as follows. In Section \ref{sec_system_model}, we present the signal and channel model of the cooperative ISAC system. In Section \ref{sec_pre_steps}, we provide several preliminary steps for the proposed cooperative sensing scheme. In Section \ref{sec_tensor_decomposition}, the parameter estimation problem is formulated via using a tensor decomposition model and a spatial smoothing tensor decomposition scheme is developed. In Section \ref{sec_coop}, we estimate the positions and velocities of the UAVs. In Section \ref{sec_dual_polar}, we extend our approach to the dual-polarized system. Finally, the simulations and conclusion are provided in Sections \ref{sec_simulation} and \ref{sec_conclusion}, respectively. 

{\itshape Notations:} Lowercase letter, boldface lowercase letter, boldface uppercase letter and calligraphy uppercase letter denote the scalars, vectors, matrices and tensors, respectively, i.e., $y$, ${\bf y}$, ${\bf Y}$, $\boldsymbol{\mathcal{Y}}$. The operations $(\cdot)^{\ast}$, $(\cdot)^{T}$, $(\cdot)^{H}$ and $(\cdot)^{\dagger}$ represent the conjugate, transpose, Hermitian transpose, and Pseudo-inverse, respectively. The notations $\mathrm{Tr}(\mathbf{Y})$, $\|\mathbf{Y}\|_{F}$, $\|\mathbf{y}\|_{2}$ and $|y|$ denote the trace of matrix $\mathbf{Y}$, the Frobenius norm of matrix $\mathbf{Y}$, the L2-norm of vector $\mathbf{y}$, and the modulus of scalar $y$, respectively. The notations $\mathbb{R}$ and $\mathbb{C}$ represent the real field and the complex field, respectively. The symbol $[\mathbf{Y}]_{ij}$ refers to the $(i,j)$-th entry of matrix $\mathbf Y$. The operators $\circ$, $\otimes $ and $\odot $ denote the outer product, Kronecker product and Khatri-Rao product, respectively. The notation $\mathrm{D}\left({\bf y}\right)$ denotes the diagonal matrix formed by vector $\bf y$. The operation $\mathrm{unvec}_{M\times N}(\mathbf{y})$ rearranges $MN\times 1$ vector $\mathbf{y}$ into $M\times N$ matrix $\mathbf{Y}$. For two sets $\mathcal{A}$ and $\mathcal{B}$, $\mathcal{A}\cup \mathcal{B}$ denotes the set $\left\{x|x\in \mathcal{A} \ \mathrm{or}\ x\in \mathcal{B}\right\}$, and $\mathcal{A}\setminus \mathcal{B}$ denotes the set $\{x|x \in \mathcal{A} \ \mathrm{and} \ x\notin  \mathcal{B}\}$.

\section{System Model}\label{sec_system_model}
\subsection{Signal Model}
\begin{figure}[t]
	\centering
	\includegraphics[width=2.25in]{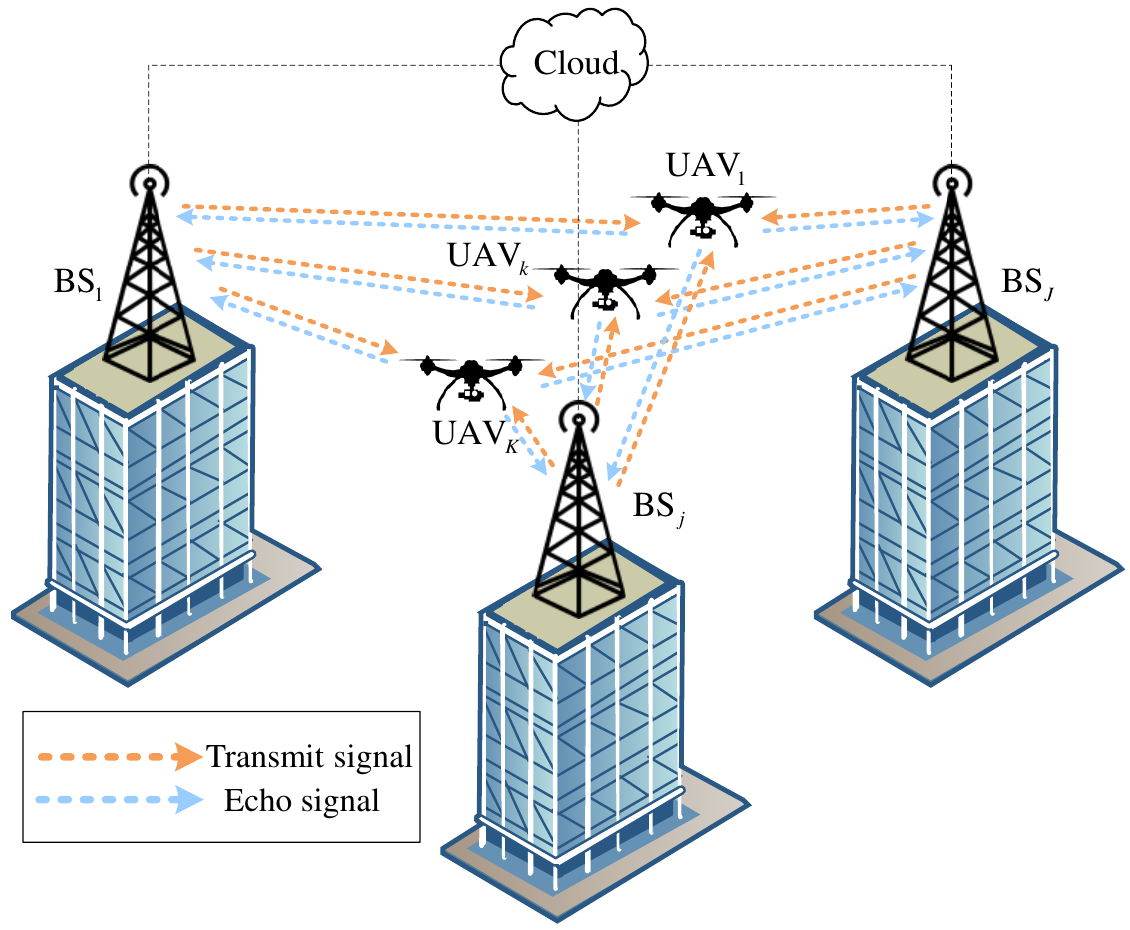}
	\caption{Cooperative ISAC system.}
	\label{fig_system_model}
\end{figure}

\begin{figure}[t]
	\centering
	\includegraphics[width=2.5in]{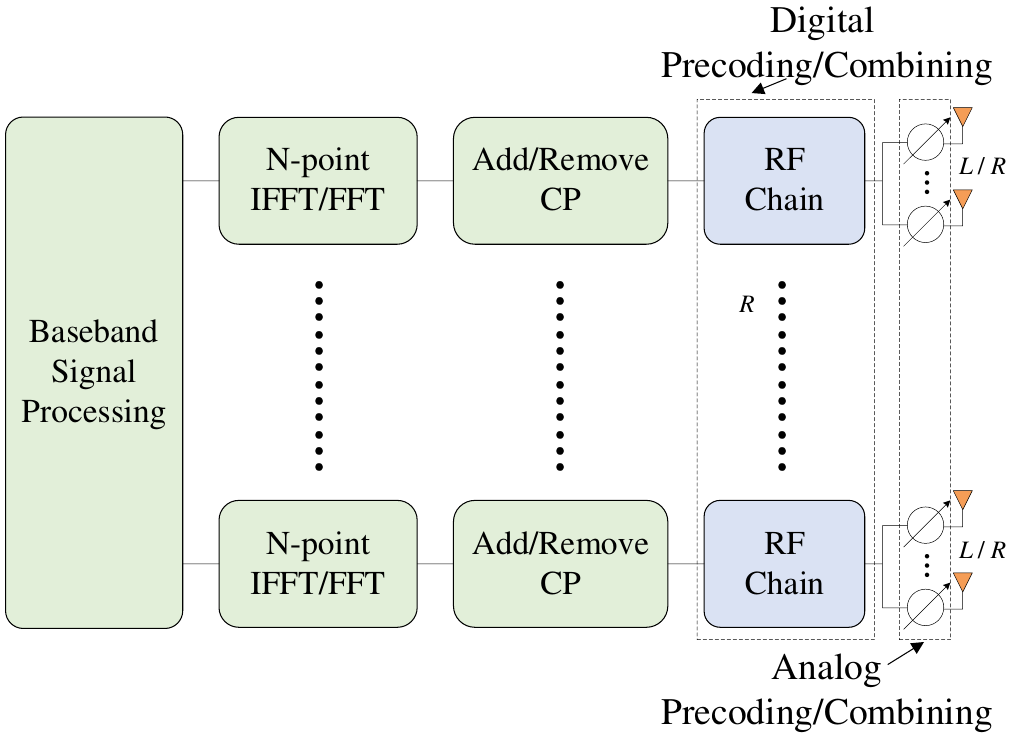}
	\caption{Partially-connected HBF structure.}
	\label{fig_HBF_structure}
\end{figure}

As shown in Fig. \ref{fig_system_model}, we consider a cooperative ISAC system aiming at sensing the low-altitude UAVs' flight status, including their positions and velocities. In this system, $J$ BSs first estimate the parameters of UAVs from the echo signals, then the estimated parameters are fused in the cloud to enhance the sensing performance. \textcolor{black}{In order to avoid the interference between the BSs, we assume that each BS operates in the non-overlapping frequency band.} In addition, as shown in Fig. \ref{fig_HBF_structure}, to achieve the high spectral efficiency and flexible resource allocation\cite{5281762,6094142,9746355} while reducing the deployment costs of radio frequency (RF) chains, we consider a MIMO-OFDM framework with partially-connected hybrid beamforming (HBF) structure, where each BS is equipped with $R$ RF chains and $L$ antennas. Specifically, each RF channel is assumed to be equipped with $L/R$ antennas, where $L/R$ is assumed to be an integer. Moreover, we assume that all the BSs perform the same parameter estimation algorithm before the data fusion is performed. Thus, for notational simplicity, we omit the subscript $j$ of BSs temporarily. In order to facilitate the formulation of parameter estimation problem via a using tensor decomposition model in the following contents of this paper, we assume that all RF channels of each BS share only one data stream for sensing. In this way, the transmitted signal is given by 
\begin{equation}
	\begin{aligned}
	&\ s\left( t \right)\\
	=&\sum_{m=0}^{M-1}{\sum_{n=0}^{N-1}{s_{m,n}\cdot e^{j2\pi m\Delta ft}\cdot r\left( t-nT_s \right), n=0,\dots ,N-1}},
	\end{aligned}
\end{equation} 
where $M$ and $N$ denote the numbers of subcarriers and OFDM symbols, respectively. The notation $\Delta f$ denotes the subcarrier spacing (SCS), $T_s$ denotes the OFDM symbol period (including the cyclic prefix), and $r\left(t\right)$ is the transmit pulse shaping filter, respectively. The symbol $s_{m,n}$ denotes the complex data allocated in the $m$-th subcarrier and the $n$-th OFDM symbol. Without loss of generality, we assume that $|s_{m,n}|^2=1,\forall m,n$. Then, the transmitted frequency domain signal vector can be expressed as
\begin{equation}
	\mathbf{x}_{m,n} =\mathbf{F}_{TX}\mathbf{e}\cdot s_{m,n} \in \mathbb{C} ^{L\times 1},
\end{equation}
where $\mathbf{F}_{TX}\triangleq\mathbf{F}_{TX}^{A}\mathbf{F}_{TX}^{D}
$, $\mathbf{F}_{TX}^{A}\in\mathbb{C}^{L\times R}$ and $\mathbf{F}_{TX}^{D}\in\mathbb{C}^{R\times R}$ denote the transmit analog and digital precoding matrices, respectively. The non-zero elements of $\mathbf{F}_{TX}^{A}$ are subjected to the constant module constraint. The symbol $\mathbf{e}=\left[1,\dots,1\right]^T\in\mathbb{R}^{R\times 1}$ denotes an all-one vector. The received frequency domain signal vector is given by 
\begin{equation}\label{E_receive_signal}
	\mathbf{y}_{m,n}={\mathbf{F}_{RX}^H}\mathbf{H}_{m,n}\mathbf{x}_{m,n} +{\mathbf{F}_{RX}^H}\mathbf{n}_{m,n}\in\mathbb{C}^{R\times 1},
\end{equation}
where $\mathbf{F}_{RX}\triangleq\mathbf{F}_{RX}^{A}\mathbf{F}_{RX}^{D}
$, $\mathbf{F}_{RX}^{A}\in\mathbb{C}^{L\times R}$ and $\mathbf{F}_{RX}^{D}\in\mathbb{C}^{R\times R}$ denote the receive analog and digital combining matrices, respectively. The non-zero elements of $\mathbf{F}_{RX}^{A}$ are also subjected to the constant module constraint. $\mathbf{H}_{m,n}$ is the discrete frequency domain sensing channel, which will be given in the following subsection. The notation $\mathbf{n}_{m,n}$ denotes the additive white Gaussian noise (AWGN). Then, we multiply the received signal vector by the conjugate of the transmitted data to eliminate its impacts, i.e.,
\begin{equation}\label{E_receive_signal_after_match}
	\tilde{\mathbf{y}}_{m,n} = s_{m,n}^{*}\mathbf{y}_{m,n}={\mathbf{F}^H_{RX}}\mathbf{H}_{m,n}\mathbf{f}_{TX} + \tilde{\mathbf{n}}_{m,n},
\end{equation}
where $\mathbf{f}_{TX} \triangleq \mathbf{F}_{TX}\mathbf{e}$, and $\tilde{\mathbf{n}}_{m,n}\triangleq s_{m,n}^{*}\mathbf{F}_{RX}^H\mathbf{n}_{m,n}$ is the equivalent noise. 

\subsection{Sensing Channel Model}\label{subsec_channel_modeling}
We assume that there are $K$ UAVs in the cooperative ISAC system. In order to derive the positions and velocities of the UAVs in 3-D space, each BS is equipped with an uniform planar array (UPA) with $P$ and $Q$ antennas located in horizontal and vertical direction,  respectively. Thus, the horizontal and  vertical steering vector can be respectively represented as 
\begin{equation}
	\mathbf{a}_p\left( \theta _k,\phi _k \right) =\left[ 1,\dots,e^{j2\pi \left( P-1 \right) d\sin(\theta_k)\cos(\phi_k) /\lambda} \right] ^T\in \mathbb{C} ^{P\times 1},
\end{equation}
\begin{equation}
	\mathbf{a}_q\left( \theta _k\right) =\left[ 1,\dots,e^{j2\pi \left( Q-1 \right) d\cos(\theta_k) /\lambda}\right] ^T\in \mathbb{C} ^{Q\times 1},
\end{equation}
where $d$ and $\lambda$ denote the antenna spacing and the wavelength, respectively. The notations $\theta_k$ and $\phi_k$ denote the elevation and azimuth angles of the $k$-th UAV, respectively. For notational simplicity, we define the virtual angles as $\vartheta\triangleq \sin(\theta)\cos(\phi)$ and $\psi\triangleq \cos(\theta)$. In this way, the steering vector of UPA can be rewritten in a compact form as 
\begin{equation}
	\mathbf{a}\left( \vartheta _k,\psi _k \right) =\mathbf{a}_q\left( \psi_k \right) \otimes \mathbf{a}_p\left( \vartheta _k \right) \in \mathbb{C} ^{PQ\times 1}.
\end{equation}
For the considered low-altitude sensing scenario, we assume that there always exists the line-of-sight (LoS) paths between the BSs and the UAVs, while the echo signals reflected by other scatterers are very weak. \textcolor{black}{In addition, we ignore the signals from other BSs (including those transmitted by other BSs and then reflected by the UAVs to the current BS). Because in the case of multi-BS frequency division configuration, the signals from other BSs can be effectively suppressed by each BS designing its own frequency band filter before receiving the echo signals.} Thus, the time and delay domain sensing channel can be expressed as
\begin{equation}
\mathbf{H}(t,\tau )=\sum_{k=1}^K{\alpha _k}\delta \left( \tau -\tau _k \right) \mathbf{a}\left( \vartheta _k,\psi _k \right) \mathbf{a}\left( \vartheta _k,\psi _k \right) ^H\cdot e^{j2\pi f_{k}^{d}t},
\end{equation}
where $\alpha_k$ denotes the channel coefficient, $\tau_k=\frac{2d_k}{c_0}$ and $f_k^d=\frac{2v_k}{\lambda}$ are the echo delay and the Doppler frequency shift caused by the $k$-th UAV, respectively. The notation $c_0$ denotes the speed of light. The notations $d_k$ and $v_k$ denote the range between the $k$-th UAV and the BS and its radial velocity to the BS, respectively. Then, by performing the Fourier transform (FT) of delay $\tau$ and sampling the received signal at the $n$-th OFDM symbol, the discrete frequency domain channel can be expressed as
\begin{equation}\label{E_frequency_domain_channel_model}
	\mathbf{H}_{m,n}\!=\!\sum_{k=1}^K{\alpha _k}\mathbf{a}\left( \vartheta _k,\psi _k \right) \mathbf{a}\left( \vartheta _k,\psi _k \right) ^H\cdot e^{-j2\pi m\Delta f\tau _k}\cdot e^{j2\pi f_{k}^{d}n T_s}.
\end{equation}

\section{\textcolor{black}{Preliminary Steps for \\The Cooperative Sensing Scheme}}\label{sec_pre_steps}	
\textcolor{black}{To enhance the practicality of the proposed cooperative sensing scheme, we present several preliminary steps to provide some prior information and guidelines for the subsequent sensing scheme.}

\subsection{\textcolor{black}{Beam Scanning and UAV Detection}}
\textcolor{black}{It should be noted that in the scenario where the UAVs are close to the BSs (e.g., less than a hundred meters), the signal-to-noise ratio (SNR) of the echo signals is sufficient to meet the requirements of UAV detection and parameter estimation. In such cases, precise localization of UAVs can be achieved directly through the parameter estimation schemes without additional beam scanning or alignment procedures. However, by considering the scenario where the UAVs may be far from the BSs, the following beam scanning and alignment are required to counteract the severe path loss, thereby improving the SNR to facilitate the subsequent parameter estimation. }

\textcolor{black}{As shown in Fig. \ref{sub_fig_beam_sweep}, we first determine the approximate locations of the UAVs via beam scanning. To further enhance the SNR of the received signals, we combine all the received symbols on each RF chain. In this way, the received signal can be expressed as
\begin{equation}
	\check{y}_{m,n}=\mathbf{e}^T\tilde{\mathbf{y}}_{m,n}=\mathbf{f}_{RX}^{H}\mathbf{H}_{m,n}\mathbf{f}_{TX}+\check{n}_{m,n},
\end{equation}
where $\mathbf{f}_{TX}$ and $\mathbf{f}_{RX}\triangleq\mathbf{F}_{RX}\mathbf{e}$ respectively represent the equivalent beamforming vector and combining vector, $\mathbf{e}=\left[1,\dots,1\right]^T\in\mathbb{R}^{R\times 1}$ denotes an all-one vector, and $\check{n}_{m,n}$ denotes the equivalent noise. During the beam scanning period, we assume that
\begin{equation}\label{E_scan_beam_design}
	\mathbf{f}_{RX}\left( \theta _i,\phi _j \right) =\mathbf{f}_{TX}\left( \theta _i,\phi _j \right) =\sqrt{\frac{P_T}{PQ}}\mathbf{a}\left( \vartheta _i,\psi _j \right), 
\end{equation}
where $P_T$ denotes the transmit power of BS, and 
\begin{subequations}
	\begin{align}
		\theta _i&=\theta _0+i\Delta \theta,\ 
		i=0,\dots ,N_{\textrm{H}}-1,
		\\
		\phi _j&=\phi _0+j\Delta \phi, \ 
		j=0,\dots ,N_{\textrm{V}}-1,
	\end{align}
\end{subequations}
where $\theta_0$ and $\phi_0$ denote the initial scan angles, $\Delta\theta$ and $\Delta\phi$ denote the scan angle steps, $i$ and $j$ denote the beam indices, $N_{\mathrm{H}}$ and $N_{\mathrm{V}}$ respectively denote the number of beams on horizontal and vertical directions, and the total number of scan beams is give by $N_{\textrm{H}}N_{\textrm{V}}$. The equivalent beamforming and combining design in (\ref{E_scan_beam_design}) can be simply achieved by setting the digital precoding/combining matrix to the identity matrix and filling the non-zero elements of the analog precoding/combining matrix with the steering vector at $(\theta_i,\phi_j)$.}

\textcolor{black}{To detect the presence of UAVs within the beam range, we formulate the following binary hypothesis testing (BHT) problem as \cite{9201513}
	\begin{equation}
		\check{y}_{m,n}=\left\{ 
		\begin{array}{l}
			\mathcal{H}_0: \check{n}_{m,n},\\
			\mathcal{H}_1: \mathbf{f}_{RX}^H \mathbf{H}_{m,n} \mathbf{f}_{TX} + \check{n}_{m,n},
		\end{array} 
		\right.
	\end{equation}
	where the null hypothesis ($\mathcal{H}_0$) assumes that the BS receives only noise, whereas the alternative hypothesis ($\mathcal{H}_1$) suggests that the BS receives both the reflected echo signals and noise. To solve the above BHT problem, we first need to construct a detector $\mathcal{T}\left(\cdot \right)$ to map $\tilde{y}_{m,n}$ to a real number, and then compare it with a predefined threshold $\gamma$ to determine whether to accept $\mathcal{H}_0$ or $\mathcal{H}_1$\cite{9201513}, i.e.,
	\begin{equation}
		\mathcal{T}\left(\check{y}_{m,n}\right) \underset{\mathcal{H}_0}{\stackrel{\mathcal{H}_1}{\gtrless}} \gamma.
	\end{equation}
	By considering the particular scenarios and available prior knowledge, various hypothesis testing methods such as likelihood ratio test (LRT) can be utilized to develop a detector\cite{Sta_signal_process}. For further manipulation, we introduce $\left\{d_{i,j}\right\}_{i=0,j=0}^{N_{\textrm{H}}-1,N_{\textrm{V}}-1}$ as the detection flag. Specifically, if the UAVs are detected within the beam range with index $(i,j)$, then $d_{i,j} = 1$; otherwise, $d_{i,j} = 0$.}

\subsection{\textcolor{black}{Estimation of the Number of UAVs}}
\textcolor{black}{For the beam range where the presence of UAVs is detected (i.e., $d_{i,j}=1$), it is necessary to further estimate the number of UAVs within the beam range, which can be achieved by considering the information theoretic criteria\cite{9724260}. Specifically, we first estimate the covariance of the received signal $\tilde{\mathbf{y}}_{m,n}$ as
\begin{equation}
	\hat{\mathbf{R}}=\frac{1}{MN}\sum_{m=0}^{M-1}{\sum_{n=0}^{N-1}{\tilde{\mathbf{y}}_{m,n}\tilde{\mathbf{y}}_{m,n}^{H}}}\in\mathbb{C}^{R\times R}.
\end{equation}
Then, we perform the singular value decomposition (SVD) of $\hat{\mathbf{R}}$, i.e., $\hat{\mathbf{R}}=\mathbf{U\Lambda U}^H$, where $\mathbf{\Lambda}=\mathrm{D}([\lambda_1,\dots,\lambda_R])$ is a diagonal matrix with the eigenvalues sorted in descending order, i.e., $\lambda _1\ge \lambda _2\ge \dots \ge \lambda_R$. Subsequently, by adopting the minimum description length (MDL) criterion\cite{1164557}, the number of UAVs within the beam range is estimated as \footnote{\textcolor{black}{Noting that we assume that the UAVs outside the current beam range will not be detected by the side lobes of the current beam. This is because the side lobe power is much less than the main lobe power. Additionally, we can further eliminate the redundant detections of the same UAVs within adjacent beams via the pruning procedures proposed in \cite{9724260}.}}
\begin{equation}
	K_{i,j}=\underset{k \in\left\{1, \ldots, R-1\right\}}{\arg \min }\mathrm{MDL}(k),
\end{equation}
with 
\begin{equation}
	\begin{aligned}
		\mathrm{MDL}(k)=&-\ln \left(\frac{\prod_{i=k+1}^{R} \lambda_i^{1 /\left(R-k\right)}}{\frac{1}{R-k} \sum_{i=k+1}^{R} \lambda_i}
		\right)^{\left(R-k\right)MN}\\
		&+\frac{1}{2}k(2R-k)\ln\left(MN\right).
	\end{aligned}
\end{equation}
}

\subsection{\textcolor{black}{Beam Alignment}}
\textcolor{black}{After the aforementioned steps, the approximate locations of the UAVs in 3-D space can be determined by recording the indices of elements that are equal to 1 in $\left\{d_{i,j}\right\}_{i=0,j=0}^{N_{\textrm{H}}-1,N_{\textrm{V}}-1}$, and the total number of UAVs can be derived by summing up the estimated number of UAVs in each beam range as
\begin{equation}
	K=\sum_{i=0,j=0}^{N_{\mathrm{H}}-1,N_{\mathrm{V}}-1}{d_{i,j}K_{i,j}},
\end{equation}
which provide valuable prior information and guidelines for the subsequent monostatic parameter estimation. Specifically, as illustrated in Fig. \ref{sub_fig_beam_alignment}, we divide the entire antenna array into $N^{\prime}=\sum_{i=0,j=0}^{N_{\mathrm{H}}-1,N_{\mathrm{V}}-1}{d_{i,j}}$ groups with each group formulating the beam aligned with the directions of $(\theta_i,\phi_j)$. For instance, we can assume that digital precoding matrix as the identity matrix, while the non-zero elements of the analog precoding matrix are filled with the normalized steering vectors corresponding to the direction of $(\theta_i,\phi_j)$. However, the non-zeros elements of the combining matrix are chosen uniformly from a normalized unit circle to mitigate the AoA estimation ambiguity caused by the dimensional reduction in the HBF structure. Noting that the aforementioned beam-related steps have been widely studied and are not the focus of this paper, in the following part of this paper, we assume that the required prior information, i.e., the beam detection flag $\left\{d_{i,j}\right\}_{i=1,j=1}^{N_{\textrm{H}},N_{\textrm{V}}}$ and the total number of the UAVs $K$, have been accurately derived via the aforementioned steps.
\begin{figure}[t]
	\centering
	\subfigure[\textcolor{black}{Beam scanning.}]{\includegraphics[width=0.14\textwidth]{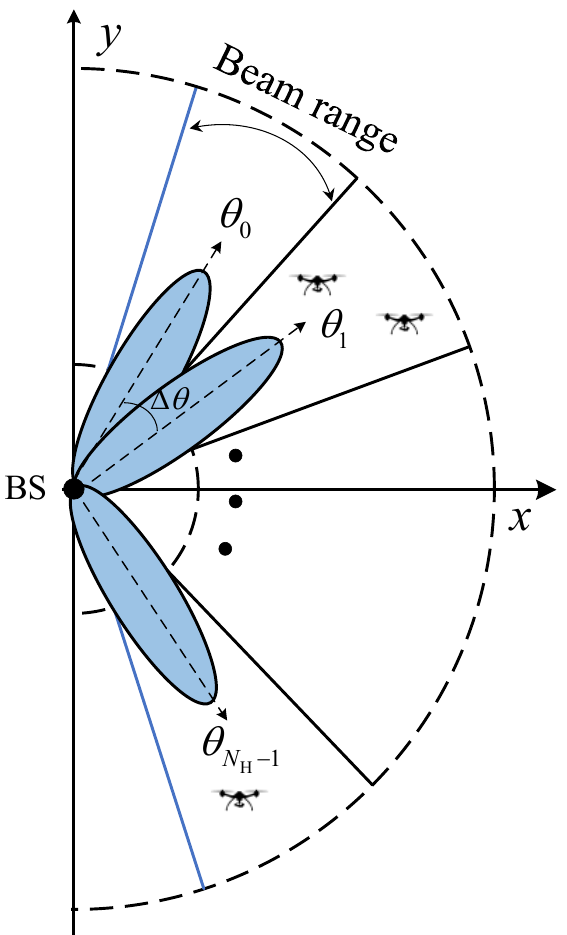}\label{sub_fig_beam_sweep}}
	\ 
	\subfigure[\textcolor{black}{Beam alignment.}]{\includegraphics[width=0.18\textwidth]{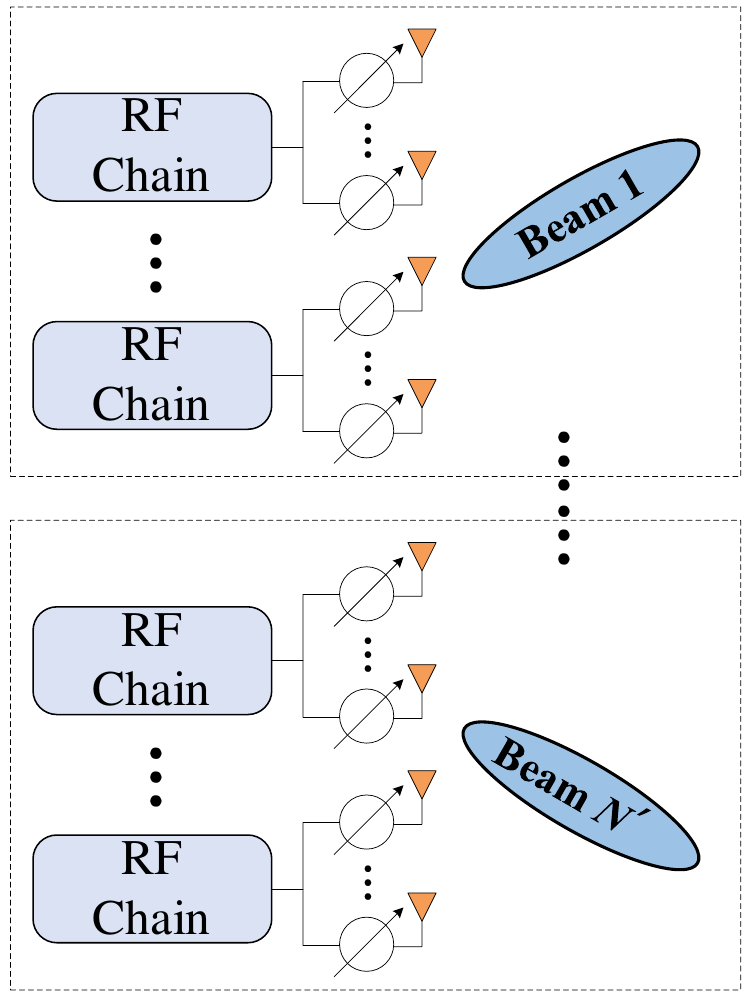}\label{sub_fig_beam_alignment}}
	\caption{\textcolor{black}{An illustration of the beam scanning and alignment.}}
\end{figure}\label{fig_beam_procedure}
}

\section{Tensor Decomposition Approach For \\Parameter Estimation}\label{sec_tensor_decomposition}
\textcolor{black}{In this section, we begin by providing several preliminaries of tensor.} Then, we formulate the parameter estimation problem via using a tensor decomposition model. Subsequently, a parameter estimation scheme based on spatial smoothing tensor decomposition is developed. Finally, we discuss the uniqueness and complexity of the tensor decomposition steps. 

\subsection{\textcolor{black}{Tensor Preliminaries}}\label{sub_sec_tensor_pre}
\textcolor{black}{To enhance the readability of this paper, some basic theory and key definitions about tensor are provided. We recommend the readers to refer to \cite{tensor} for more details.}

\textcolor{black}{{\itshape{1) Unfolding:}} Mode-$n$ unfolding of a tensor $\boldsymbol{\mathcal{X}}\in\mathbb{C}^{I_1\times I_2 \times \dots \times I_N}$ denotes that one rearranges the tensor to a matrix. Specifically, tensor element $\left(i_1, i_2, \dots, i_N\right)$ maps to matrix element $\left(i_n, j\right)$, where  
\begin{equation}
	j=1+\sum_{\substack{k=1 \\ k \neq n}}^N\left(i_k-1\right) J_k,\ \text { with } \  J_k=\prod_{\substack{m=1 \\ m \neq n}}^{k-1} I_m.
\end{equation}}

\textcolor{black}{{\itshape{2) Rank-1 Tensor:}} An $N$-th-order tensor $\boldsymbol{\mathcal{X}}$ is a rank-1 tensor if it can be expressed as the outer product of $N$ vectors, i.e.,
\begin{equation}
	\boldsymbol{\mathcal{X}} =\mathbf{a}^{(1)}\circ \mathbf{a}^{(2)}\circ \cdots \circ \mathbf{a}^{\left( N \right)}. 
\end{equation}}

\textcolor{black}{{\itshape{3) CP decomposition:}}
CANDECOMP/PARAFAC (CP) decomposition denotes that one decomposes a tensor into a sum of component rank-1 tensors, i.e.,
\begin{equation}
	\boldsymbol{\mathcal{X}} =\sum_{r=1}^R{\lambda _r}\mathbf{a}_{r}^{(1)}\circ \mathbf{a}_{r}^{(2)}\circ \cdots \circ \mathbf{a}_{r}^{(N)},
\end{equation}
where $R$ denotes the rank of $\boldsymbol{\mathcal{X}}$. The corresponding factor matrix to the $n$-th mode is defined as $\mathbf{A}^{\left( n \right)}=\left[ \mathbf{a}_{1}^{(n)},\dots ,\mathbf{a}_{R}^{(n)} \right] ,n=1,\dots ,N$.
The mode-$n$ unfolding version of $\boldsymbol{\mathcal{X}}$ is given by
\begin{equation}
	\begin{aligned}
		&\mathbf{X}_{(n)}=\\
		&\mathbf{A}^{(n)}\mathbf{\Lambda }\left( \mathbf{A}^{(N)}\odot \cdots \odot \mathbf{A}^{(n+1)}\odot \mathbf{A}^{(n-1)}\odot \cdots \odot \mathbf{A}^{(1)} \right) ^T,
	\end{aligned}
\end{equation}
where $\mathbf{\Lambda }=\mathrm{D}\left([\lambda_1,\dots,\lambda_R]\right)$.}

\subsection{Tensor Formulation}
Recalling the expression of the received signal and the channel model provided in Section \ref{sec_system_model}, \textcolor{black}{we find that the AoAs, Doppler shifts and time delays caused by the UAVs independently affect the received signal in spatial, time and frequency dimensions, which prompts us to formulate the multi-dimensional parameter estimation via using a tensor decomposition model.} Specifically, by substituting (\ref{E_frequency_domain_channel_model}) into (\ref{E_receive_signal_after_match}) and stacking the received signal among $N$ OFDM symbols and $M$ subcarriers, it is readily verified that the received signal vector $\tilde{\mathbf{y}}_{m,n}$ can be formulated into a third-order tensor as
\begin{equation}
	\boldsymbol{\mathcal{Y}} =\sum_{k=1}^K{\alpha _k\mathbf{b}\left( \vartheta _k,\psi _k \right) \circ \mathbf{o}\left( f_{k}^{d} \right) \circ}\mathbf{g}\left( \tau _k \right) +\boldsymbol{\mathcal{N}} \in \mathbb{C} ^{R\times N\times M},
\end{equation}
where $\boldsymbol{\mathcal{N}}\in\mathbb{C}^{R\times N \times M}$ is the equivalent noise tensor, \textcolor{black}{$K$ denotes the total number of UAVs, which can be derived by the preliminary steps provided in Section \ref{sec_pre_steps}}, and
\begin{equation}
	\mathbf{b}\left( \vartheta _k,\psi _k \right) ={\mathbf{F}^H_{RX}}\mathbf{a}\left( \vartheta _k,\psi _k \right) \mathbf{a}\left( \vartheta _k,\psi _k \right) ^H\mathbf{f}_{TX}\in \mathbb{C} ^{R\times 1},
\end{equation}
\begin{equation}
	\mathbf{o}\left( f_{k}^{d} \right) =\left[ 1,e^{j2\pi T_sf_{k}^{d}},...,e^{j2\pi \left( N-1 \right) T_sf_{k}^{d}} \right] ^T\in \mathbb{C} ^{N\times 1},
\end{equation}
\begin{equation}
	\mathbf{g}\left( \tau _k \right) =\left[ 1,e^{-j2\pi \Delta f\tau _k},...,e^{-j2\pi \left( M-1 \right) \Delta f\tau _k} \right] ^T\in \mathbb{C} ^{M\times 1},
\end{equation}
respectively. The corresponding factor matrices of $\boldsymbol{\mathcal{Y}}$ are given by
\begin{equation}
	\mathbf{A}^{\left( 1 \right)}=\left[ \mathbf{b}(\vartheta _1,\psi _1),\dots ,\mathbf{b}(\vartheta _K,\psi _K) \right] \in \mathbb{C} ^{R\times K},
\end{equation}
\begin{equation}
	\mathbf{A}^{\left( 2 \right)}=\left[\mathbf{o}(f_{1}^{d}),\dots ,\mathbf{o}(f_{K}^{d}) \right] \in \mathbb{C} ^{N\times K},
\end{equation}
\begin{equation}
	\mathbf{A}^{\left( 3 \right)}=\left[ \alpha_1\mathbf{g}(\tau _1),\dots ,\alpha_K\mathbf{g}(\tau _K) \right] \in \mathbb{C} ^{M\times K}.
\end{equation}
Then, the tensor decomposition model is given by
\begin{equation}\label{E_TALS}
	\min_{\mathbf{A}^{(1)},\mathbf{A}^{(2)},\mathbf{A}^{(3)}} \left\| \boldsymbol{\mathcal{Y}} -\sum_{k=1}^K{\alpha _k\mathbf{b}\left( \vartheta _k,\psi _k \right) \circ \mathbf{o}\left( f_{k}^{d} \right) \circ}\mathbf{g}\left( \tau _k \right) \right\| _{F}^{2}.
\end{equation}

\subsection{Factor Matrices Recovery}
Instead of solving Problem (\ref{E_TALS}) via the well-known alternative least square (ALS) method\cite{7914672}, we recover the factor matrices by leveraging the Vandermonde structure of $\mathbf{A}^{(3)}$. For notational simplicity, the noise component is ignored in the following derivations. First, the mode-1 unfolding of the third-order tensor $\boldsymbol{\mathcal{Y}}$ can be expressed as\cite{tensor}
\begin{equation}\label{E_Mode1_unfolding}
	\mathbf{Y}_{\left( 1 \right)}^T=\left( \mathbf{A}^{\left( 3 \right)}\odot \mathbf{A}^{\left( 2 \right)} \right)
	\left(\mathbf{A}^{\left(1\right)}\right)^T\in \mathbb{C} ^{MN\times R}.
\end{equation}
Then, we choose a pair of integer $\{L_1,L_2\}$ satisfying $L_1+L_2=M+1$ and define the following cyclic choose matrix as
\begin{equation}\label{E_cyclic_choose_matrix}
	\mathbf{J}_l=\left[ \mathbf{0}_{L_1\times (l-1)},\mathbf{I}_{L_1},\mathbf{0}_{L_1\times \left( L_2-l \right)} \right] \in \mathbb{C} ^{L_1\times M}.
\end{equation}
We then smooth the mode-1 unfolding of $\boldsymbol{\mathcal{Y}}$ as
\begin{equation}\label{E_smooth_Y1}
	\begin{aligned}
		\mathbf{Y}^S&=\left[ \left( \mathbf{J}_1\otimes \mathbf{I}_N \right) \mathbf{Y}_{(1)}^{T},\dots ,\left( \mathbf{J}_{L_2}\otimes \mathbf{I}_N \right) \mathbf{Y}_{(1)}^{T} \right]\\
		&\overset{(a)}{=}\left( \mathbf{A}^{(L_1,3)}\odot \mathbf{A}^{\left( 2 \right)} \right) \left( \mathbf{A}^{(L_2,3)}\odot \mathbf{A}^{\left( 1 \right)} \right) ^T \in \mathbb{C} ^{L_1N\times L_2R},
	\end{aligned}
\end{equation}
where $\mathbf{A}^{(L,3)}$ denotes the 1 to $L$ rows of $\mathbf{A}^{(3)}$. Equation $(a)$ is derived by leveraging the property of Khatri-Rao product, i.e., $\left( \mathbf{A}\otimes \mathbf{B} \right) \left( \mathbf{C}\odot \mathbf{D} \right) =\left( \mathbf{AC} \right) \odot \left( \mathbf{BD} \right)$\cite{zxd} and the Vandermonde structure of $\mathbf{A}^{(3)}$, while the details are omitted due to the space limitations. We then perform the truncated SVD of $\mathbf{Y}_S$ as
\begin{equation}\label{E_SVD_of_YS}
	\mathbf{Y}^S=\mathbf{U\Sigma V}^H,
\end{equation}
where $\mathbf{U}\in \mathbb{C} ^{L_1 N\times K}$, $\mathbf{\Sigma }\in \mathbb{C} ^{K\times K}$ and $\mathbf{V}\in \mathbb{C} ^{L_2 R\times K}$. 
Given that the columns of $\bf{U}$ span the same subspace as the columns of $\mathbf{Y}^S$, there always exists a full rank matrix $\mathbf{M}\in \mathbb{C} ^{K\times K}$ satisfying\cite{6573422}
\begin{equation}\label{E_A3_KR_A2}
	\mathbf{A}^{\left( L_1,3 \right)}\odot \mathbf{A}^{\left( 2 \right)}=\mathbf{UM}\in \mathbb{C} ^{L_1N\times K},
\end{equation}
\begin{equation}\label{E_A3_KR_A1}
	\mathbf{A}^{\left( L_2,3 \right)}\odot \mathbf{A}^{\left( 1 \right)}=\mathbf{V}^*\mathbf{\Sigma M}^{-T}\in \mathbb{C} ^{L_2R\times K}.
\end{equation}
Noting the Vandermonde structure of $\mathbf{A}^{\left( L_1,3 \right)}$, we have 
\begin{equation}\label{E_A3_Vandermonde}
	\left( \underline{\mathbf{A}}^{\left( L_1,3 \right)}\odot \mathbf{A}^{\left( 2 \right)} \right) \mathbf{Z}=\overline{\mathbf{A}}^{(L_1,3)}\odot \mathbf{A}^{\left( 2 \right)},
\end{equation}
where $\mathbf{Z}=\mathrm{D}\left([e^{-j2\pi \Delta f\tau _1},\dots ,e^{-j2\pi \Delta f\tau _K}]\right)$, $\underline{\mathbf{A}}^{\left( L_1,3 \right)}$ and $\overline{\mathbf{A}}^{\left( L_1,3 \right)}$ denote the deletions of the last row and the first row of $\mathbf{A}^{\left( L_1,3 \right)}$, respectively. Then, one obtains
\begin{equation}\label{E_A3_underline_KR_A2}
	\underline{\mathbf{A}}^{\left( L_1,3 \right)}\odot \mathbf{A}^{\left( 2 \right)}=\mathbf{U}_1\mathbf{M},
\end{equation}
\begin{equation}\label{E_A3_overline_KR_A2}
	\overline{\mathbf{A}}^{(L_1,3)}\odot \mathbf{A}^{\left( 2 \right)}=\mathbf{U}_2\mathbf{M}, 
\end{equation}
where $\mathbf{U}_1=\mathbf{U}_{1:\left( L_1-1 \right) N,:}$ denotes the 1 to $\left( L_1-1 \right) N$ rows of $\mathbf{U}$, and $\mathbf{U}_2=\mathbf{U}_{ N+1:L_1 N,:}$ denotes the $N+1$ to $L_1 N$ rows of $\mathbf{U}$, respectively. Then, combining (\ref{E_A3_underline_KR_A2}), (\ref{E_A3_overline_KR_A2}) with (\ref{E_A3_Vandermonde}), we have
\begin{equation}\label{E_EVD_of_pinv_U1_U2}
	\mathbf{U}_1\mathbf{MZ}=\mathbf{U}_2\mathbf{M}\Rightarrow \mathbf{MZM}^{-1}=\mathbf{U}_1^{\dagger}\mathbf{U}_2\triangleq\mathbf{\Xi}.
\end{equation}
Thus, by performing the eigenvalue decomposition (EVD) of $\mathbf{\Xi}$, we can derive the estimations of $\mathbf{Z}$ and $\mathbf{M}$. Since the normalized diagonal elements of $\mathbf{Z}$, i.e., $\hat{z}_k=\left[\mathbf{Z}\right]_{k,k}/|\left[\mathbf{Z}\right]_{k,k}| , k=1,\dots, K$ are actually the generators of $\mathbf{A}^{\left(3\right)}$, each column of $\mathbf{A}^{\left(3\right)}$ can be recovered as
\begin{equation}
	\hat{\mathbf{a}}_k^{(3)}=\left[1,\hat{z}_k,\dots,\hat{z}_k^{M-1}\right]^T,k=1,\dots, K.
\end{equation}
Then, recalling the definition of Khatri-Rao product, we have
\begin{equation}\label{E_KR_defination}
	\mathbf{A}^{\left( L_1,3 \right)}\odot \mathbf{A}^{\left( 2 \right)}=\left[ \mathbf{a}_{1}^{(L_1,3)}\otimes \mathbf{a}_{1}^{\left( 2 \right)},\dots ,\mathbf{a}_{K}^{(L_1,3)}\otimes \mathbf{a}_{K}^{\left( 2 \right)} \right] =\mathbf{UM}.
\end{equation}
\textcolor{black}{Thus, given $\hat{\mathbf{A}}^{\left(L_1,3\right)}$ and $\mathbf{M}$, each column of $\mathbf{A}^{\left(2\right)}$ can be derived as\cite{6573422}
\begin{equation}\label{E_each_column_of_A2}
	\begin{aligned}
	\mathbf{a}_{k}^{\left( 2 \right)}&\overset{\left( a \right)}{=}\left( \frac{{\mathbf{a}_{k}^{(L_1,3)}}^H}{\left\| \mathbf{a}_{k}^{(L_1,3)} \right\| _{2}^{2}}\otimes \mathbf{I}_N \right) \left( \mathbf{a}_{k}^{(L_1,3)}\otimes \mathbf{a}_{k}^{(2)} \right)\\ 
	&\overset{\left( b \right)}{=}\left( \frac{{\mathbf{a}_{k}^{(L_1,3)}}^H}{\left\| \mathbf{a}_{k}^{(L_1,3)} \right\| _{2}^{2}}\otimes \mathbf{I}_N \right) \mathbf{Um}_k,k=1,\dots ,K,
\end{aligned}
\end{equation}
where $\left(a\right)$ is obtained by using the mixed-producted property of Kronecker product, i.e., $(\mathbf{A} \otimes \mathbf{B})(\mathbf{C} \otimes \mathbf{D})=\mathbf{A} \mathbf{C} \otimes \mathbf{B D}$, $\left(b\right)$ is based on (\ref{E_KR_defination}), and $\mathbf{m}_k$ denotes the $k$-th column of $\mathbf{M}$. Similarly, letting $\mathbf{P}\triangleq\mathbf{M}^{-T}$, we can recover each column of $\mathbf{A}^{\left(1\right)}$ as \cite{6573422}
\begin{equation}\label{E_each_column_of_A1}
\begin{aligned}
	\mathbf{a}_{k}^{\left( 1 \right)}&\overset{\left( a \right)}{=}\left( \frac{{\mathbf{a}_{k}^{(L_2,3)}}^H}{\left\| \mathbf{a}_{k}^{(L_2,3)} \right\| _{2}^{2}}\otimes \mathbf{I}_R \right) \left( \mathbf{a}_{k}^{(L_2,3)}\otimes \mathbf{a}_{k}^{(1)} \right) \\
	&\overset{\left( b \right)}{=}\left( \frac{{\mathbf{a}_{k}^{(L_2,3)}}^H}{\left\| \mathbf{a}_{k}^{(L_2,3)} \right\| _{2}^{2}}\otimes \mathbf{I}_R \right) \mathbf{V}^*\mathbf{\Sigma p}_k,k=1,\dots ,K,
\end{aligned}
\end{equation}
where $\mathbf{p}_k$ denotes the $k$-th column of $\mathbf{P}$.}

\subsection{Parameter Estimation}\label{Subsec_parameters_est}
In this subsection, we estimate the parameters from the recovered factor matrices. First, \textcolor{black}{the estimation of the time delay $\{{\tau}_k\}_{k=1}^{K}$ can be derived from the generators of $\mathbf{A}^{(3)}$, i.e.,}
\begin{equation}\label{E_hat_tau}
	\hat{\tau}_k=\frac{\measuredangle \hat{z}_k}{-2\pi \Delta f},
\end{equation}
where $\measuredangle{\hat{z}_k}$ denotes the angle of $\hat{z}_k$. \textcolor{black}{The estimation of the range $\{d_k\}_{k=1}^K$ between the UAVs and the BS can be derived as}
\begin{equation}\label{E_hat_d}
	\hat{d}_k=\frac{\hat{\tau}_kc_0}{2}.
\end{equation}
\textcolor{black}{The estimation of the Doppler frequency shift $\{{f}_k^d\}_{k=1}^{K}$ can be derived by the following correlation-based scheme as\cite{7914672}}
\begin{equation}\label{E_hat_fd}
	\hat{f}_{k}^{d}=\mathrm{arg}\max_{f_{k}^{d}} \left| \hat{\mathbf{o}}_{k}^{H}\mathbf{o}\left( f_{k}^{d} \right) \right|^2,
\end{equation}
where $\hat{\mathbf{o}}_{k}$ denotes the $k$-th column of $\hat{\mathbf{A}}^{\left(2\right)}$. \textcolor{black}{The radial velocity $\{v_k\}_{k=1}^K$ of the UAVs to the BS can be estimated as}
\begin{equation}\label{E_hat_radial_velocity}
	\hat{v}_k=\frac{\hat{f}_{k}^{d}\lambda}{2}.
\end{equation} 
\textcolor{black}{Similarly, the estimation of $\{{\vartheta}_k, {\psi}_k\}_{k=1}^K$ can be derived as}
\begin{equation}\label{E_2D_search}
	\{ \hat{\vartheta}_k,\hat{\psi}_k\} =\mathrm{arg}\max_{\vartheta,\psi} \frac{\left| \hat{\mathbf{b}}_{k}^{H}\mathbf{b}(\vartheta,\psi) \right|^2}{\left\| \mathbf{b}(\vartheta,\psi) \right\|_2 ^2},
\end{equation}
where $\hat{\mathbf{b}}_{k}$ denotes the $k$-th column of $\hat{\mathbf{A}}^{\left(1\right)}$. Noting that Problem (\ref{E_hat_fd}) can be directly solved by the one-dimensional (1-D) search method. However, performing the two-dimensional (2-D) search method in (\ref{E_2D_search}) will result in a heavy computational burden. \textcolor{black}{Thus, we develop a low-complexity AoA estimation algorithm based on GRQ to address Problem (\ref{E_2D_search}).} Specifically, the objective function (OF) of Problem (\ref{E_2D_search}) can be rewritten as (\ref{E_2D_OF_reformulation}) shown at the bottom of this page, where equation $\left(a\right)$ is derived by eliminating the common factor $\mathbf{a}\left( \vartheta ,\psi \right) ^H\mathbf{f}_{TX}\mathbf{f}_{TX}^{H}\mathbf{a}\left( \vartheta ,\psi \right)$ from both the numerator and denominator, equation $\left(b\right)$ comes from the reformulation of  $\mathbf{a}\left(\vartheta,\psi\right)$. \textcolor{black}{In specific, we have
\begin{equation}
	\begin{aligned}
		{\bf{a}}\left( {\theta ,\psi } \right) &= \left( {{{\bf{a}}_q}\left( \psi  \right) \cdot 1} \right) \otimes \left( {{{\bf{I}}_p}{{\bf{a}}_p}\left( \theta  \right)} \right) \\
		&= \left( {{{\bf{a}}_q}\left( \psi  \right) \otimes {{\bf{I}}_p}} \right){{\bf{a}}_p}\left( \theta  \right), 
	\end{aligned}
\end{equation}
which can be derived by leveraging the mixed-producted property of Kronecker product, i.e., $\left( \mathbf{A}\otimes \mathbf{B} \right) \left( \mathbf{C}\otimes \mathbf{D} \right) =\mathbf{AC}\otimes \mathbf{BD}$.} Equation $\left(c\right)$ is obtained by defining 
\begin{subequations}
	\begin{align}
		\mathbf{Q}_1^k(\psi )\triangleq&\left[ \mathbf{a}_q(\psi )\otimes \mathbf{I}_P \right] ^H\mathbf{F}_{RX}\hat{\mathbf{b}}_k\hat{\mathbf{b}}_{k}^{H}{\mathbf{F}_{RX}^H}\left[ \mathbf{a}_q(\psi )\otimes \mathbf{I}_P \right],\label{eq_Q1}\\
		\mathbf{Q}_2^k\left( \psi \right) \triangleq&\left[ \mathbf{a}_q(\psi )\otimes \mathbf{I}_P \right] ^H\mathbf{F}_{RX}{\mathbf{F}_{RX}^H}\left[ \mathbf{a}_q(\psi )\otimes \mathbf{I}_P \right],
	\end{align}
\end{subequations}
respectively.
\begin{figure*}[hb] 
	\centering
	\hrulefill 
	\vspace*{0pt} 
	\begin{equation}\label{E_2D_OF_reformulation}
		\begin{aligned}
			f\left( \vartheta ,\psi \right) &=\frac{\hat{\mathbf{b}}_{k}^{H}\mathbf{F}_{RX}^{H}\mathbf{a}\left( \vartheta ,\psi \right) \mathbf{a}\left( \vartheta ,\psi \right) ^H\mathbf{f}_{TX}\mathbf{f}_{TX}^{H}\mathbf{a}\left( \vartheta ,\psi \right) \mathbf{a}\left( \vartheta ,\psi \right) ^H\mathbf{F}_{RX}\hat{\mathbf{b}}_k}{\mathbf{f}_{TX}^{H}\mathbf{a}\left( \vartheta ,\psi \right) \mathbf{a}\left( \vartheta ,\psi \right) ^H\mathbf{F}_{RX}\mathbf{F}_{RX}^{H}\mathbf{a}\left( \vartheta ,\psi \right) \mathbf{a}\left( \vartheta ,\psi \right) ^H\mathbf{f}_{TX}}\overset{(a)}{=}\frac{\mathbf{a}\left( \vartheta ,\psi \right) ^H\mathbf{F}_{RX}\hat{\mathbf{b}}_k\hat{\mathbf{b}}_{k}^{H}\mathbf{F}_{RX}^{H}\mathbf{a}\left( \vartheta ,\psi \right)}{\mathbf{a}\left( \vartheta ,\psi \right) ^H\mathbf{F}_{RX}\mathbf{F}_{RX}^{H}\mathbf{a}\left( \vartheta ,\psi \right)}
			\\
			&\overset{(b)}{=}\frac{\mathbf{a}_p(\vartheta )^H\left[ \mathbf{a}_q(\psi )\otimes \mathbf{I}_P \right] ^H\mathbf{F}_{RX}\hat{\mathbf{b}}_k\hat{\mathbf{b}}_{k}^{H}\mathbf{F}_{RX}^{H}\left[ \mathbf{a}_q(\psi )\otimes \mathbf{I}_P \right] \mathbf{a}_p(\vartheta )}{\mathbf{a}_p(\vartheta )^H\left[ \mathbf{a}_q(\psi )\otimes \mathbf{I}_P \right] ^H\mathbf{F}_{RX}\mathbf{F}_{RX}^{H}\left[ \mathbf{a}_q(\psi )\otimes \mathbf{I}_P \right] \mathbf{a}_p(\vartheta )}\overset{(c)}{=}\frac{\mathbf{a}_p(\vartheta )^H\mathbf{Q}_{1}^{k}(\psi )\mathbf{a}_p(\vartheta )}{\mathbf{a}_p(\vartheta )^H\mathbf{Q}_{2}^{k}(\psi )\mathbf{a}_p(\vartheta )}.
		\end{aligned}
	\end{equation}
	\hrulefill
\end{figure*}
Then, Problem (\ref{E_2D_search}) can be reformulated as
\begin{equation}\label{E_Rayleigh_quotient}
	\underset{\vartheta ,\psi}{\max}\quad \frac{\mathbf{a}_p(\vartheta )^H\mathbf{Q}_1^k(\psi )\mathbf{a}_p(\vartheta )}{\mathbf{a}_p(\vartheta )^H\mathbf{Q}_2^k(\psi )\mathbf{a}_p(\vartheta )}.
\end{equation}
Noting that with fixed $\psi$, Problem (\ref{E_Rayleigh_quotient}) has a well-known GRQ form\cite{zxd}. It is readily verified that $\mathbf{a}_p(\vartheta )$ is an eigenvector of the matrix $\mathbf{\Phi }^k\left( \psi \right) \triangleq \left( \mathbf{Q}_{2}^{k}(\psi ) \right) ^{\dagger}\mathbf{Q}_{1}^{k}(\psi )$, and the OF value of Problem (\ref{E_Rayleigh_quotient}) is the corresponding eigenvalue of $\mathbf{\Phi }^k\left( \psi \right)$. Consequently, we estimate ${\psi}_k$ as
\begin{equation}\label{E_1D_search_for_psi}
	\hat{\psi}_k=\mathrm{arg}\max_{\psi} \  \lambda _{\max}\{ 
	\mathbf{\Phi }^k\left( \psi \right)\}, 
\end{equation}
where $\lambda _{\max}\{\cdot\}$ denotes the maximum eigenvalue of a matrix, and the above problem can be solved by the 1-D search method.
At the first glance, performing the EVD of $\mathbf{\Phi }^k\left( \psi \right)$ to obtain the maximum eigenvalue of $\mathbf{\Phi }^k\left( \psi \right)$ on several grids will still incur considerable calculation burden. However, the following lemma proves that the EVD of $\mathbf{\Phi }^k\left( \psi \right)$ on each grid is unnecessary. 

\itshape \textbf{Lemma 1}: \upshape Problem (\ref{E_1D_search_for_psi}) is equivalent to the following problem:
\begin{equation}\label{E_1D_search_for_psi_by_trace}
	\hat{\psi}_k=\mathrm{arg}\max_{\psi} \,\,\mathrm{Tr}\left(\mathbf{\Phi }^k\left( \psi \right)
	\right).
\end{equation}

\itshape \textbf{Proof:} \upshape According to the expression of (\ref{eq_Q1}), it is readily verified that the rank of $\mathbf{Q}_1^k\left(\psi\right)$ is 1. Then, based on the property of the rank of matrix multiplication \cite{zxd}, i.e., 
\begin{equation}
	\mathrm{rank}\left( \mathbf{AB} \right) \le \min \left\{ \mathrm{rank}\left( \mathbf{A} \right) ,\mathrm{rank}\left( \mathbf{B} \right) \right\}, 
\end{equation}
we can prove that the rank of $\mathbf{\Phi }^k\left( \psi \right)$ is always 1. In addition, according to the property of rank-1 matrix\cite{zxd}, we have 
\begin{equation}
	\mathrm{Tr}\left(\mathbf{\Phi }^k\left( \psi \right)\right) =\lambda,
\end{equation}
where $\lambda$ denotes the unique non-zero eigenvalue of $\mathbf{\Phi }^k\left( \psi \right)
$. Additionally, it is readily verified that $\mathbf{\Phi }^k\left( \psi \right)$ is a positive semi-definite matrix, thus, we have $
\mathrm{Tr}\left( \mathbf{\Phi }^k\left( \psi \right) \right)=\lambda =\lambda _{\max}\left\{\mathbf{\Phi }^k\left( \psi \right)\right\}$ and the proof is completed. \hfill $\blacksquare$

According to Lemma 1, we address Problem (\ref{E_1D_search_for_psi_by_trace}) to estimate $\psi _k$ by the 1-D search method. Subsequently, the estimation of ${\mathbf{a}}_p\left(\vartheta\right)$ is given by the eigenvector corresponding to the maximum eigenvalue of $\mathbf{\Phi }^k(\hat{\psi}_k)$. We let $
\hat{\mathbf{a}}_p(\vartheta )\gets \hat{\mathbf{a}}_p(\vartheta )/[\hat{\mathbf{a}}_p(\vartheta )]_1$ to eliminate the scaling ambiguity brought by the EVD, where $[\cdot]_1$ denotes the first element of a vector. Then, we estimate ${\vartheta}_k$ as
\begin{equation}\label{E_1D_search_for_vartheta}
	\hat{\vartheta}_k=\mathrm{arg}\min_{\vartheta}\ \left\| \mathbf{a}_p\left( \vartheta \right) -\hat{\mathbf{a}}_p\left( \vartheta \right) \right\| _2,
\end{equation}
which can still be solved by the 1-D search method. The detailed procedures of the proposed GRQ-based algorithm are summarized in Algorithm \ref{Algo_RD_GRB}. Then, the estimation of the elevation and azimuth angles can be derived as
\begin{equation}\label{E_hat_theta}
	\hat{\theta}_k=\mathrm{arc} \cos(\hat{\psi}_k),
\end{equation}
\begin{equation}\label{E_hat_phi}
	\hat{\phi}_k=\mathrm{arc} \cos\left(\frac{\hat{\vartheta}_k}{\sin(\hat{\theta}_k)}\right).
\end{equation}
Finally, we eliminate the scaling ambiguity and estimate the channel coefficients as \cite{9049103}
\begin{subequations}
	\begin{align}
		\left[ \mathbf{\Lambda }_1 \right] _{k,k}&=\mathbf{b}^{\dagger}\left( \hat{\vartheta}_k,\hat{\psi}_k \right) \hat{\mathbf{b}}_k,\label{E_Gamma_1}
		\\
		\left[ \mathbf{\Lambda }_2 \right] _{k,k}&=\mathbf{o}^{\dagger}\left( \hat{f}_{k}^{d} \right) \hat{\mathbf{o}}_k,
		\label{E_Gamma_2}
		\\
		\mathbf{\Lambda }_3&=\left({\mathbf{\Lambda }_1}\right)^{-1}\left({\mathbf{\Lambda }_2}\right)^{-1},\label{E_Gamma_3}
		\\
		\hat{\alpha}_k&=\left( \left[ \mathbf{\Lambda }_3 \right] _{k,k}\mathbf{g}\left( \hat{\tau}_{k} \right) \right) ^{\dagger}\hat{\mathbf{g}}_k.\label{E_hat_alpha}
	\end{align}
\end{subequations}
The detailed procedures of the tensor-based parameter estimation scheme are summarized in Algorithm \ref{Algo_Tensor_Decomposition}.

\begin{algorithm}[t]
	\caption{\textcolor{black}{GRQ-Based Method to Solve Problem} (\ref{E_2D_search})}\label{Algo_RD_GRB}
	\begin{algorithmic}[1]
		\STATE 
		Estimate ${\psi}_k$ in  (\ref{E_1D_search_for_psi_by_trace}) by the 1-D search method;
		\STATE
		Calculate the maximum eigenvalue of $\mathbf{\Phi }^k(\hat{\psi}_k )$ and the corresponding eigenvector $\hat{\mathbf{a}}_p(\vartheta)$;
		\STATE
		Calculate $\hat{\mathbf{a}}_p(\vartheta )\gets \hat{\mathbf{a}}_p(\vartheta)/[\hat{\mathbf{a}}_p(\vartheta)]_1$ to eliminate the scaling ambiguity;
		\STATE
		Estimate ${\vartheta}_k$ in (\ref{E_1D_search_for_vartheta}) by the 1-D search method. 
	\end{algorithmic}
\end{algorithm}

\subsection{Uniqueness Analysis}\label{subsec_uniqueness_analysis}
In this subsection, we discuss the uniqueness of the above tensor decomposition, which guarantees the correct recovery of the factor matrices and the subsequent parameter estimation. A well-known Kruskal sufficient uniqueness condition is provided in \cite{Kruskal}. In addition, by leveraging more structural information of factor matrices, the uniqueness can be further relaxed as \cite{6573422,10403776}

\itshape \textbf{Lemma 2}: \upshape Let $\boldsymbol{\mathcal{X}} \in \mathbb{C}^{I_1 \times I_2 \times I_3}$
be a tensor with three factor matrices $\mathbf{A}^{\left(1\right)}\in\mathbb{C}^{I_1 \times K}$, $\mathbf{A}^{\left(2\right)}\in\mathbb{C}^{I_2 \times K}$ and $\mathbf{A}^{\left(3\right)}\in\mathbb{C}^{I_3 \times K}$, where $\mathbf{A}^{\left(3\right)}$ is a Vandermonde matrix with distinct generators $\{z_k\}_{k=1}^K$. Denote $k_{\mathbf{A}}$ as the Kruskal-rank of matrix $\mathbf{A}$, if
\begin{equation}\label{E_Uniqueness_condition_with_Vandermone}
	\left\{\begin{array}{l}
		k_{\left(\underline{\mathbf{A}}^{(L_1, 3)} \odot \mathbf{A}^{(2)}\right)}=K,\\
		k_{\left(\mathbf{A}^{(L_2, 3)}\odot \mathbf{A}^{(1)}\right)}=K,
	\end{array}\right.
\end{equation}
\noindent then the rank of $\boldsymbol{\mathcal{X}}$ is $K$, and the tensor decomposition is unique. In the generic case, condition (\ref{E_Uniqueness_condition_with_Vandermone}) becomes
\begin{equation}
	\min \left(\left(L_1-1\right) I_2, L_2 I_1\right) \geq K.
\end{equation}
 
\itshape \textbf{Proof:} \upshape Please refer to\cite{6573422}.\hfill $\blacksquare$

Recalling the formulation of $\boldsymbol{\mathcal{Y}}$, the notations $I_1$, $I_2$ and $I_3$ in Lemma 2 respectively denote the number of RF chains $R$, the number of OFDM symbols $N$, and the number of subcarriers $M$. Since there are a large number of subcarriers in current MIMO-OFDM systems, we can readily choose a pair of integer $\{L_1,L_2\}$ satisfying $L_1+L_2 = M+1$, $\left(L_1-1\right)I_2 = (L_1-1)N\ge K$ and $L_2 I_1 = L_2 R\ge K$. Therefore, the uniqueness condition can be readily satisfied. \textcolor{black}{In addition, the uniqueness implies that the AoAs, Doppler shifts, and time delays will automatically be associated with the same target without designing additional pairing procedures\cite{6573422}.}

\begin{algorithm}[t]
	\caption{\textcolor{black}{Spatial Smoothing Tensor Decomposition Approach For Parameter Estimation}}\label{Algo_Tensor_Decomposition}
	\begin{algorithmic}[1]
		\STATE 
		Mode-1 unfold $\boldsymbol{\mathcal{Y}}$ as (\ref{E_Mode1_unfolding});
		\STATE
		Choose $\{L_1, L_2\}$ and smooth $\mathbf{Y}_{(1)}^T$ to $\mathbf{Y}^S$ as (\ref{E_smooth_Y1});
		\STATE
		Perform the SVD of $\mathbf{Y}^S$ as (\ref{E_SVD_of_YS});
		\STATE
		Perform the EVD of $\mathbf{\Xi}$ as (\ref{E_EVD_of_pinv_U1_U2});
		\STATE
		Calculate the generators $\hat{z}_k=[\mathbf{Z}]_{k,k}/|[\mathbf{Z}]_{k,k}|, K=1,\dots, K$;
		\STATE
		Construct each column of $\mathbf{A}^{\left(3\right)}$ by
		$\hat{\mathbf{g}}(\tau_k)=\left[1,\hat{z}_k,\dots,\hat{z}_k^{M-1}\right]^T,k=1,\dots, K$;
		\STATE
		Construct each column of $\mathbf{A}^{\left(2\right)}$ via (\ref{E_each_column_of_A2});
		\STATE
		Construct each column of $\mathbf{A}^{(1)}$ via (\ref{E_each_column_of_A1});
		\STATE
		\textcolor{black}{Derive the time delay $\{\hat{\tau}_k\}_{k=1}^K$, range $\{\hat{d}_k\}_{k=1}^K$, Doppler frequency shift $\{\hat{f}_k^d\}_{k=1}^K$, radial velocity $\{\hat{v}_k\}_{k=1}^K$,  elevation angle $\{\hat{\theta}_k\}_{k=1}^K$ and azimuth angle $\{\hat{\phi}_k\}_{k=1}^K$ via (\ref{E_hat_tau}), (\ref{E_hat_d}), (\ref{E_hat_fd}), (\ref{E_hat_radial_velocity}), (\ref{E_hat_theta}) and (\ref{E_hat_phi}), respectively. }
		
		\STATE
		Eliminate the scaling ambiguity and estimate the channel coefficients via (\ref{E_Gamma_1}), (\ref{E_Gamma_2}), (\ref{E_Gamma_3}) and (\ref{E_hat_alpha}), respectively. 
	\end{algorithmic}
\end{algorithm}

\subsection{Complexity Analysis}
We next analyze the complexities of the steps in Algorithm 2. It should be pointed out that the complexity of step 2 is negligible, since we can leverage the property of the sparse cyclic choose matrices, i.e., $\mathbf{J}_{l}$ to collect specific rows of $\mathbf{Y}_{(1)}^{T}$ and stack them into the extended matrix $\mathbf{Y}^{S}$ instead of actually multiplying $\mathbf{Y}_{(1)}^{T}$ by $\mathbf{J}_l$. At step 3, we perform the truncated SVD of $\mathbf{Y}^{S}$, which has the complexity of $\mathcal{O}\left(L_1 N K L_2 R\right)$. At step 4, performing the EVD of $\mathbf{\Xi}$ incurs the complexity of $\mathcal{O}\left(K^3\right)$. From step 5 to step 6, the reconstruction of $\mathbf{A}^{\left(3\right)}$ mainly requires the complexity of $\mathcal{O}\left(KM\right)$. Next, the total complexity of the reconstruction of $\mathbf{A}^{\left(2\right)}$ and $\mathbf{A}^{\left(1\right)}$ is given by $\mathcal{O}\left(L_1 N^2 K + L_1 N K^2 + L_2 R^2 K + L_2 R K^2\right)$. In the stage of parameter estimation, the derivations of $\{\hat{f}^d_k\}_{k=1}^K$ has the complexity of $\mathcal{O}\left(NGK\right)$, and the complexity of Algorithm 1 is given by $\mathcal{O}\left(P^3GK+PGK\right)$, where $G$ denotes the grid size. Finally, step 10 mainly has the complexity of $\mathcal{O}\left(\left(R+M+N\right)K\right)$. Thus, the total complexity of Algorithm 2 can be expressed as $\mathcal{O}\big{(}L_1 N KL_2 R + K^3 + KM + L_1 N^2K + L_1NK^2 + L_2 R^2 K + L_2 R K^2 + NGK +P^3GK\big{)}$.

\section{Position and Velocity Estimation}\label{sec_coop}
\textcolor{black}{In this section, we enhance the positioning accuracy of the UAVs and recover their true velocities through the multi-BS cooperation, as the above monostatic parameter estimation method can only derive the UAVs' radial velocities.} Specifically, we first develop a false removing MST-based data association method, and then estimate the positions and velocities of the multiple UAVs relying on the parameters estimated by the BSs, i.e., $\{\hat{\theta}_{k,j},\hat{\phi}_{k,j},\hat{d}_{k,j},\hat{v}_{k,j}\}_{k=1,j=1}^{K,J}$, where the subscript $j$ denotes the BS's index. 

\subsection{False Removing MST-Based Data Association}
\begin{figure}[t]
	\centering
	\includegraphics[width=1.8in]{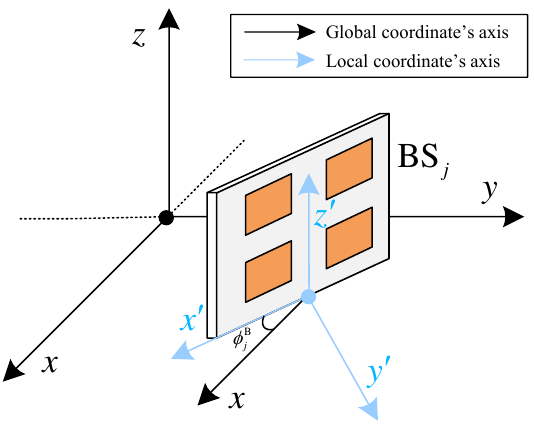}
	\caption{The relationship between the local and the global coordinate.}
	\label{fig_Panel}
\end{figure}
Noting that in conventional device-based sensing for active targets that can send/receive RF signals with different signatures, the BSs know the exact mapping between the estimation parameters of a certain target, i.e., each BS knows the true indices for all $K$ UAVs. However, in the considered device-free sensing scenario, the detected UAVs' indices may be different across BSs. Thus, before performing the data fusion, we develop a false removing MST-based data association method. The initial phase of data association involves each BS roughly estimating the positions of $K$ UAVs relying on the AoA and range estimations, i.e., $\{\hat{\theta}_{k,j},\hat{\phi}_{k,j},\hat{d}_{k,j}\}_{k=1,j=1}^{K,J}$. Specifically, the $k$-th UAV's position estimated by $\textrm{BS}_j$ is given by
\begin{equation}\label{E_hat_p_kj}
	\hat{\mathbf{p}}_{k,j}=\hat{d}_{k,j}\hat{\mathbf{r}}_{k,j}+\mathbf{p}_j^{\textrm{B}},
\end{equation}
where $\mathbf{p}_j^{\textrm{B}}$ denotes the position of $\text{BS}_j$ in the global coordinate, and 
\begin{equation}\label{E_dir_vector}
\hat{\mathbf{r}}_{k,j}=\mathbf{T}\left( \phi _{j}^{\mathrm{B}} \right) \left[ \begin{array}{c}
	\sin \left( \hat{\theta}_{k,j} \right) \cos \left( \hat{\phi}_{k,j} \right)\\
	\sin \left( \hat{\theta}_{k,j} \right) \sin \left( \hat{\phi}_{k,j} \right)\\
	\cos \left( \hat{\theta}_{k,j} \right)\\
\end{array} \right], 
\end{equation}
is the direction vector in the global coordinate. The notation $\mathbf{T}\left( \phi _{j}^{\mathrm{B}}\right) $ denotes the transform matrix from the $\text{BS}_j$'s local coordinate to the global coordinate. Without loss of generality, as shown in Fig. \ref{fig_Panel}, we assume that the antenna panel of each BS is parallel to the $z$-axis of the global coordinate, and $\phi_j^{\textrm{B}}$ is the angle between the $x$-axis of the $\textrm{BS}_j$'s local coordinate and the $x$-axis of the global coordinate. In this way, the transform matrix can be expressed as \cite{han2024cellular}
\begin{equation}
	\mathbf{T}\left( \phi _{j}^{\mathrm{B}}\right) =\left[ \begin{matrix}
		\cos \left( \phi _{j}^{\mathrm{B}} \right)&		\sin \left( \phi _{j}^{\mathrm{B}} \right)&		0\\
		-\sin \left( \phi _{j}^{\mathrm{B}} \right)&		\cos \left( \phi _{j}^{\mathrm{B}} \right)&		0\\
		0&		0&		1\\
	\end{matrix} \right]. 
\end{equation}
Then, according to the positioning results of BSs, i.e., $\{\hat{\mathbf{p}}_{k,j}\}_{k=1,j=1}^{K,J}$ and leveraging the similarity of positioning results for the same UAV across multiple BSs, a straightforward association approach is to minimize the sum of the Euclidean distances between the positioning results from different BSs via the exhaustive permutation method\cite{9583869}. However, the complexity rises sharply with the value of $K$. In addition, the permutation method does not remove the false detections, while it is evident that integrating these false detection results with the correct detection results will greatly impact the following data fusion process.
\begin{figure}[t]
	\centering
	\includegraphics[width=2.1in]{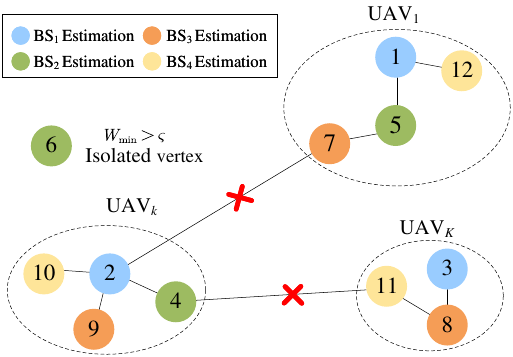}
	\caption{An illustration of the MST of graph $\tilde{G}$.}
	\label{fig_MST}
\end{figure}

To address these issues, we propose a false removing MST-based data association method. Specifically, we first construct an undirected weighted graph $G = (V, E)$. The vertex set $V$ represents the union of a series of sub-vertex sets, i.e.,
\begin{equation}
	V=V_1\cup V_2\cup \dots \cup V_J, 
\end{equation}
where $V_j=\left\{ \left( j-1 \right) K+1,\dots ,jK \right\} ,j=1,\dots ,J$ denotes the $K$ positioning results of $\textrm{BS}_j$. The edge set $E$ is defined as
\begin{equation}
	E=\left\{ \left( a,b \right) |a\in V_j,b\in V\setminus V_j,j=1,\dots ,J \right\},  
\end{equation}
and the weights of edges are defined as the Euclidean distances between the positioning results derived by BSs, i.e.,  
\begin{equation}
	\begin{aligned}
		W_{ab}=|| &\hat{\mathbf{p}}_{(a-1)\ \!\mathrm{mod}\ \! K+1,\lfloor \frac{a-1}{K} \rfloor +1}\\
		&-\hat{\mathbf{p}}_{(b-1)\ \!\mathrm{mod}\ \!K+1,\lfloor \frac{b-1}{K} \rfloor +1} {||}_2,\forall a,b\in E,
	\end{aligned}
\end{equation}
where the notations $\mathrm{mod}$ and $\lfloor \cdot \rfloor$ denote the modulus and floor operations, respectively. To prevent the false detection results from impacting the subsequent data fusion, we define the following edge set as
\begin{equation}
	E^{\prime}=\left\{ \left( a,b \right) |W_{\min}\left( a \right) >\varsigma ,a,b\in E \right\},
\end{equation}
where $W_{\min}\left( a \right)>\varsigma$ indicates that the shortest adjacent edge of vertex $a$ is longer than the threshold $\varsigma$. Then, we update the edge set as\footnote{Assuming that $\textrm{BS}_j$
successfully detects the $k$-th UAV, there always exists at least one another BS whose positioning result for the same UAV is similar to the $\textrm{BS}_j$'s estimation, guaranteeing $W_{\min}\left( a \right)\le\varsigma$, where $a$ denotes the vertex corresponding to $\textrm{BS}_j$'s estimation. In contrast, $W_{\min}\left( a \right)>\varsigma$ indicates that $\textrm{BS}_j$ fails to detect the $k$-th UAV. Thus, to prevent the false detection from impacting the following data fusion, the vertex $a$ should be recognized as an isolated vertex in the updated graph $\tilde{G}$.} 
\begin{equation}\label{E_Remove_edges}
	\tilde{E}=E\setminus E^{\prime}.
\end{equation}
In this way, in the updated graph $\tilde{G}=(V,\tilde{E})$, these false detection results are displayed as the isolated vertices. Then, we derive the MST of the connected components of graph $\tilde{G}$ via the well-known Prim\cite{6773228} or Kruskal \cite{kruskal1956shortest} algorithm. As shown in Fig. \ref{fig_MST}, given that the positioning results of multiple BSs for the same UAV are more similar than those for distinct UAVs, the MST algorithm will always connect the positioning results of multiple BSs for the same UAV. Subsequently, by removing $K-1$ longest edges from the MST, the positioning results from multiple BSs for $K$ UAVs are divided into $K$ sub-graphs. By collecting the vertex indices in each sub-graph, the data association can be immediately accomplished. In the following contents of this paper, we assume that each BS holds the same indices of UAVs. In addition, for a certain UAV, these false detections and estimations derived by BSs will not be incorporated in the data fusion, even though we still utilize the notation $J$ for notational simplicity.
\begin{figure}[t]
	\centering
	\subfigure[Rough position estimation relying on mean fusion.]{\includegraphics[width=0.22\textwidth]{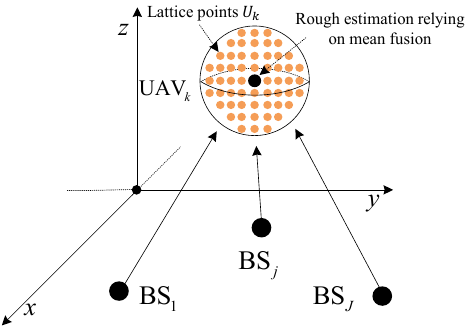}}
	\ 
	\subfigure[Calculate the OF values.]{\includegraphics[width=0.22\textwidth]{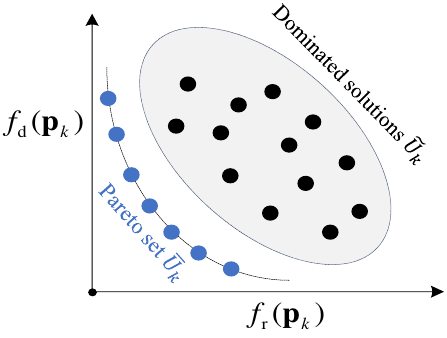}}
	\caption{An illustration of the Pareto optimality fusion scheme.}
	\label{fig_Pareto}
\end{figure}

\subsection{Position Estimation}
After accurate data association, we then perform the data fusion across the multiple BSs. In general, there are two primary fusion strategies: hard fusion and soft fusion\cite{han2024cellular}. Specifically, the former denotes that each BS estimates and sends the final positioning results, i.e., $\{\hat{\mathbf{p}}_{k,j}\}_{k=1,j=1}^{K,J}$ to the cloud. Then, the cloud integrates the position estimations via the (weighted) mean fusion scheme. For instance, the position estimation relying on mean fusion is given by
\begin{equation}\label{E_Mean_fusion}
	\hat{\mathbf{p}}_k=\frac{1}{J}\sum_{j=1}^J{\hat{\mathbf{p}}_{k,j}}, \ k=1,\dots ,K.
\end{equation}
Differently, with soft fusion strategy, all BSs send the initial parameter estimations, including the AoAs and ranges to the cloud. Then, the cloud performs the estimation of positions via integrating these uploaded parameters. Since soft fusion provides more fine-grained information for the data fusion, we mainly focus on this fusion strategy in the following contents of this paper. Specifically, noting that the accuracy of position estimation mainly depends on the range and radial direction (determined by AoAs) estimations, we respectively define the following range loss function and direction loss function as
\begin{subequations}
	\begin{align}
	f_{\mathrm{r}}\left( \mathbf{p}_k \right) =&\frac{\sum_{j=1}^J{\alpha _{k,j}\left| d_{k,j}\left( \mathbf{p}_k \right) -\hat{d}_{k,j} \right|}}{\sum_{j=1}^J{\alpha _{k,j}}},\label{E_dir_loss_func}\\
	f_{\mathrm{d}}\left( \mathbf{p}_k \right) =&\frac{\sum_{j=1}^J{\alpha _{k,j}\left\| \mathbf{r}_{k,j}\left( \mathbf{p}_k \right) -\hat{\mathbf{r}}_{k,j} \right\| _2}}{\sum_{j=1}^J{\alpha _{k,j}}},\label{E_range_loss_func}
	\end{align}
\end{subequations}
where $d_{k,j}(\mathbf{p}_k)=\left\|\mathbf{p}_k-\mathbf{p}_{j}^{\mathrm{B}} \right\| _2$ denotes the true range between the UAV and the BS, and $\mathbf{r}_{k,j}\left( \mathbf{p}_k \right)$ denotes the true radial direction vector determined by the $k$-th UAVs' true position $\mathbf{p}_k$. The notation $\alpha_{k,j}$ denotes the weight assigned to $\text{BS}_{j}$. Specifically, due to path loss, the SNR of echo signals decrease with the increase of the range between the UAV and BSs. Thus, we define the weighting coefficient as
\begin{equation}
	\alpha _{k,j}=\frac{1}{\left( d_{k,j}(\mathbf{p}_k) \right) ^{\beta _1}},
\end{equation}
 where $\beta_1>0$ is a factor introduced to control the weighting intensity. Then, we address the following problem to estimate the position of the $k$-th UAV, i.e.,
\begin{equation}\label{E_intial_loss_func}
	\underset{\mathbf{p}_k}{\min}\left\{ f_{\textrm{r}}\left( \mathbf{p}_k \right) ,f_{\textrm{d}}\left( \mathbf{p}_k \right) \right\} ,
\end{equation}
which is a highly nonlinear multi-objective optimization problem with the OFs holding the distinct dimensions. To address Problem (\ref{E_intial_loss_func}), we propose a Pareto optimality strategy to determine the UAVs' positions. Specifically, as shown in Fig. \ref{fig_Pareto}, we first take the mean fusion result in (\ref{E_Mean_fusion}) as a rough position estimation. Then, we construct $L$ lattice points around the rough estimation and calculate the two OF values corresponding to these lattice points. To minimize both two OF values simultaneously, we first recognize the dominated solutions among the lattice points, i.e.,
\begin{equation}
	\begin{aligned}
		\tilde{U}_k=\{&\mathbf{p}_{k}^{l_1}|f_{\textrm{r}}( \mathbf{p}_{k}^{l_1} ) >f_{\textrm{r}}( \mathbf{p}_{k}^{l_2} ) ,f_{\textrm{d}}( \mathbf{p}_{k}^{l_1}) >f_{\textrm{d}}( \mathbf{p}_{k}^{l_2}) ,\\
		&\mathbf{p}_{k}^{l_1},\mathbf{p}_{k}^{l_2}\in U_k\},
	\end{aligned}
\end{equation}
where $U_k=\left\{ \mathbf{p}_{k}^{1},\dots ,\mathbf{p}_{k}^{L} \right\} $ denotes the set formed by $L$ lattice points. Then, we remove the dominated solutions and retain the Pareto set of solutions as
\begin{equation}\label{E_Remove_dominated_sol}
\begin{aligned}
	\bar{U}_k=U_k\backslash\tilde{U}_k.
\end{aligned}
\end{equation}
Finally, we determine the solution from the Pareto set as the final position estimation relying on the system configuration. Specifically, in scenarios where each BS is allocated a substantial number of subcarriers but only equipped with relatively a small number of antennas, the range estimation always exhibits a higher precision than AoA estimation. Thus, the range estimation should be more dominant for positioning. In such cases, we can choose the solution from the Pareto set, which has the minimum range loss function value, i.e.,
\begin{equation}\label{E_range_dominant}
	\hat{\mathbf{p}}_k=\mathrm{arg} \underset{\mathbf{p}_{k}^{l}\in \bar{U}_k}{\min}f_{\mathrm{r}}\left( \mathbf{p}_{k}^{l} \right).
\end{equation}
On the flip side, when the BSs are equipped with a large number of antennas but a small number of subcarriers, the AoA estimation becomes the critical factor, guiding the choice of the solution with minimum direction loss function value from the Pareto set, i.e., 
\begin{equation}\label{E_AoA_dominant}
	\hat{\mathbf{p}}_k=\mathrm{arg} \underset{\mathbf{p}_{k}^{l}\in \bar{U}_k}{\min}f_{\mathrm{d}}\left( \mathbf{p}_{k}^{l} \right).
\end{equation}
In addition, we can also take into account the accuracy of AoA and range estimation in the designed monostatic parameter estimation algorithm to achieve better positioning performance. Noting that only several simple calculations are required to calculate the OF values corresponding to the lattice points, so the complexity is tolerant.

\subsection{Velocity Estimation}
\begin{figure}[t]
		\centering
		\includegraphics[width=1.8in]{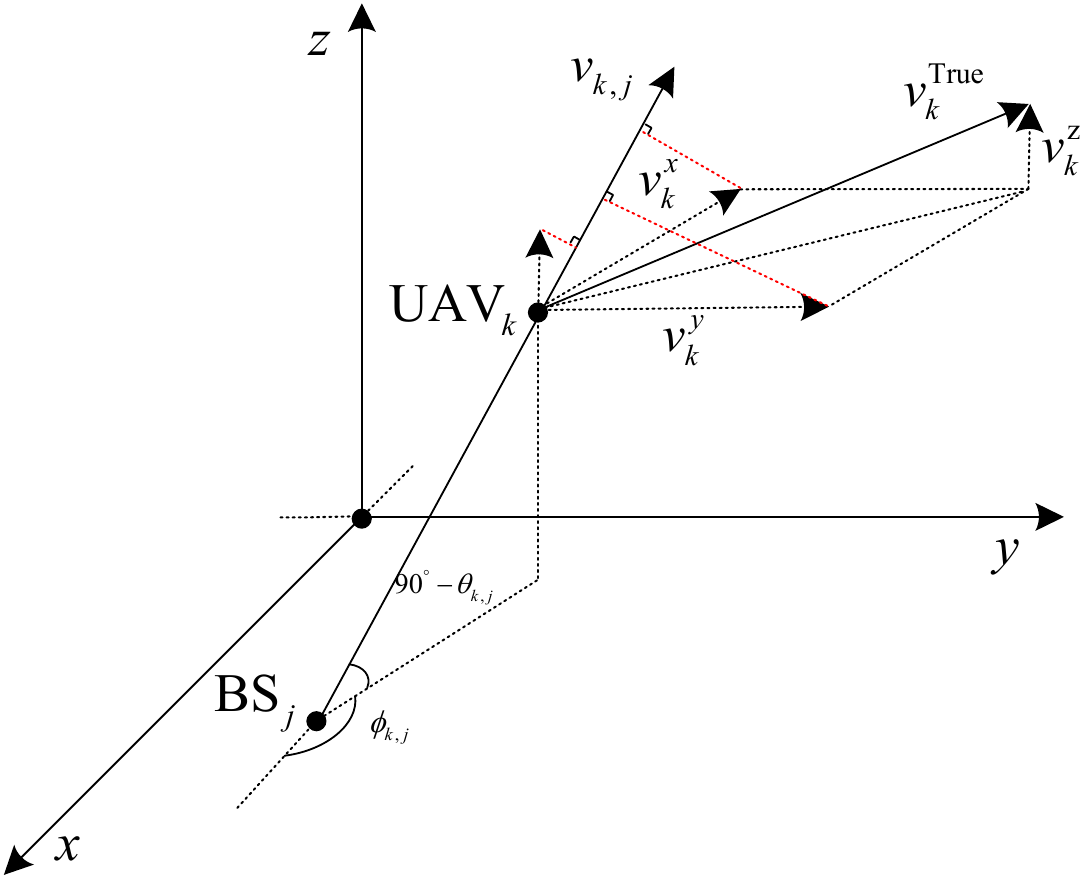}
		\caption{The relationship between the radial velocity and the true velocity (assume that the global coordinate coincides with the local coordinate).}
		\label{fig_Cal_Velocity}
\end{figure}
Noting that relying on the above positioning scheme, we always achieve an enhanced estimation of UAVs' positions via cooperation compared to the monostatic position estimation. Thus, we can calibrate the previous AoA and range estimations, i.e., $\{\hat{\theta}_{k,j},\hat{\phi}_{k,j},\hat{d}_{k,j}\}_{k=1,j=1}^{K,J}$relying on the cooperative position estimations. For notational simplicity, we continue to use the original notations to denote the calibrated estimations. 

According to the tensor decomposition procedures in Section \ref{sec_tensor_decomposition}, each BS can only estimate the radial velocities of UAVs, i.e., $\{\hat{v}_{k,j}\}_{k=1,j=1}^{K,J}$.
In this subsection, we estimate the true velocity of each UAV based on $\{\hat{v}_{k,j}\}_{k=1,j=1}^{K,J}$ and $\{\hat{\theta}_{k,j},\hat{\phi}_{k,j}\}_{k=1,j=1}^{K,J}$. To perform the estimation, we first explore the relationship between the true velocity and the radial velocity. As shown in Fig. \ref{fig_Cal_Velocity}, we denote $\mathbf{v}_{k}^{\mathrm{True}}=\left[ v_{k}^{x},v_{k}^{y},v_{k}^{z} \right] ^T$ as the true velocity of the $k$-th UAV, where the elements represent the velocity components on the three coordinate axises of the global coordinate. Then, by projecting the three components to the radial direction, we have
\begin{equation}\label{E_Relation_between_radial_and_true_velocity_in_vec_form}
	{\hat{\mathbf{r}}_{k,j}^T}\mathbf{v}_{k}^{\mathrm{True}}=\hat{v}_{k,j}+\varepsilon _{k,j}
	,\quad j=1,\dots, J, 
\end{equation}
where $\hat{\mathbf{r}}_{k,j}$ denotes the (calibrated) radial direction vector defined by (\ref{E_dir_vector}), and $\varepsilon_{k,j}$ denotes the estimation error. As such, stacking all the radial velocity estimations derived by $J$ BSs, (\ref{E_Relation_between_radial_and_true_velocity_in_vec_form}) can be rewritten in a more compact form as
\begin{equation}
	\hat{\mathbf{\Omega}}_k\mathbf{v}_{k}^{\mathrm{True}}=\hat{\mathbf{v}}_k+\boldsymbol{\varepsilon }_k,
\end{equation} 
where
\begin{equation}
	\hat{\mathbf{\Omega}}_k=\left[ \begin{array}{c}
	{\hat{\mathbf{r}}_{k,1}}^T\\
	\vdots\\
	{\hat{\mathbf{r}}_{k,J}}^T\\
	\end{array} \right] ,\hat{\mathbf{v}}_k=\left[ \begin{array}{c}
	\hat{v}_{k,1}\\
	\vdots\\
	\hat{v}_{k,J}\\
	\end{array} \right] ,\boldsymbol{\varepsilon }_k=\left[ \begin{array}{c}
	\varepsilon _{k,1}\\
	\vdots\\
	\varepsilon _{k,J}\\
	\end{array} \right]. 
\end{equation}
Then, the true velocity of the $k$-th UAV can be derived by the well-known weighted LS (WLS) estimation, i.e.,
\begin{equation}
	\hat{\mathbf{v}}_{k}^{\mathrm{True}}=\left( \hat{\mathbf{\Omega}}_{k}^{T}\mathbf{W}_k\hat{\mathbf{\Omega}}_k \right) ^{-1}\hat{\mathbf{\Omega}}_{k}^{T}\mathbf{W}_k\hat{\mathbf{v}}_k,
\end{equation}
where $\mathbf{W}_k=\mathrm{D}\left( \left[ \gamma _{k,1},\dots ,\gamma _{k,J} \right] \right)$ denotes the weighting matrix with the diagonal element $\gamma _{k,j}=\frac{1}{\left( \hat{d}_{k,j} \right) ^{\beta _2}}$ denoting the weight assigned to $\textrm{BS}_j$'s estimation, and $\beta_2>0$ denotes the weighting intensity factor. However, noting that the number of BSs in the cooperative ISAC systems is limited, which makes the above WLS method still sensitive to the estimation error. Therefore, we propose a more robust residual weighting-based method to suppress the impacts of estimation error. Specifically, the method mainly includes the following three steps\cite{797838}: 

{\itshape 1) Grouping BSs:} Noting that to recover the true velocity, at least three BSs' estimations are required. Therefore, we select all combinations from $J\ge 3$ BSs including at least three BSs. In this way, the total number of combinations is given by
\begin{equation}
I=\sum_{j=3}^J{C_{J}^{j}}.
\end{equation}
Then, we collect all the combinations into a set as $
\chi =\left\{ X_i|i=1,\dots ,I \right\}$. 

{\itshape 2) Calculating residuals:} For a certain combination, we adopt the mentioned WLS method to derive the rough estimation of the $k$-th UAV's true velocity relying on the BSs' estimations within the combination, which is denoted as $\hat{\mathbf{v}}_k^i$. Then, we calculate the residuals among all BSs, i.e.,
\begin{equation}\label{E_cal_residuals}
	\mathrm{Res}\left( X_i \right) =\sum_{j=1}^J{\gamma_{k,j}\left( \hat{\mathbf{r}}_{k,j}^{T}\hat{\mathbf{v}}_k^i-\hat{v}_{k,j} \right) ^2},i = 1,\dots, I.
\end{equation}

{\itshape 3) Weighting the estimations based on the residuals:} Relying on the calculated residuals, the true velocity is estimated as
\begin{equation}\label{E_Weighting_Residuals}
	\hat{\mathbf{v}}_{k}^{\mathrm{True}}=\frac{\sum_{i=1}^I{\left\{ \mathrm{Res}\left( X_i \right) ^{-1}\hat{\mathbf{v}}_k^i \right\}}}{\sum_{i=1}^I{\mathrm{Res}\left( X_i \right) ^{-1}}}.
\end{equation}
As previously stated, the number of BSs in cooperative ISAC systems is often limited, so the computational complexity of the above method is tolerant. When the number of BSs is large, we can adopt several existing greedy strategies to further reduce the complexity\cite{4376962}.

\textcolor{black}{Based on the above discussions, the detailed procedures of the overall cooperative position and velocity estimation method are summarized in Algorithm 3. }

\begin{algorithm}[t]
	\caption{\textcolor{black}{Cooperative Position and Velocity Estimation Scheme}}\label{Algo_Coop_Sensing}
	\begin{algorithmic}[1]
	   {\color{black}
		\STATE
		\textbf{Multi-BS data association:}
		\STATE 
		Calculate the UAVs' positions $\{\hat{\bf{p}}_{k,j}\}_{k=1,j=1}^{K,J}$ via (\ref{E_hat_p_kj});
		\STATE
		Construct graph $G$ with $\{\hat{\bf{p}}_{k,j}\}_{k=1,j=1}^{K,J}$;
		\STATE
		Update graph $\tilde{G}$ by removing the false detection results via (\ref{E_Remove_edges});
		\STATE
		Derive the MST of graph $\tilde{G}$;
		\STATE
		Remove the $K-1$ longest edges from the MST and collect the vertex indices in each sub-graph.
		\STATE 
		\textbf{Position estimation:}
		\STATE
		Derive the rough position estimation $\{\hat{\mathbf{p}}_k\}_{k=1}^K$ via (\ref{E_Mean_fusion});		
		\STATE
		Construct $L$ lattice points $U_k$ around the rough position estimation;
		\STATE
		Calculate the two OF values corresponding to lattice points $U_k$ via (\ref{E_dir_loss_func}) and (\ref{E_range_loss_func});
		\STATE
		Remove the dominated solutions and retain the Pareto set $\bar{U}_k$ via (\ref{E_Remove_dominated_sol}); 
		\STATE
		Estimate the UAVs' positions via (\ref{E_range_dominant}) or (\ref{E_AoA_dominant}).
		\STATE 
		\textbf{Velocity estimation:}
		\STATE
		Calibrate the previous AoA and range estimations, i.e., $\{\hat{\theta}_{k,j},\hat{\phi}_{k,j},\hat{d}_{k,j}\}_{k=1,j=1}^{K,J}$relying on the cooperative position estimations;
		\STATE
		Group the BSs and adopt the WLS method to derive the rough estimation of UAVs' true velocity;
		\STATE
		Calculate the residuals of each combination of BSs via (\ref{E_cal_residuals});
		\STATE
		Weight the estimations with residuals via (\ref{E_Weighting_Residuals}).}
	\end{algorithmic}
\end{algorithm}

\section{Extension to The Dual-Polarized System}\label{sec_dual_polar}
In this section, we extend the proposed tensor decomposition parameter estimation scheme to the dual-polarized system. Specifically, the sensing channel matrix is modified as \cite{8481590}
\begin{equation}\label{E_Dual_Polarized_Channel}
\mathbf{H}_{m,n}=\left[ \begin{matrix}
	\mathbf{H}_{m,n}^{(\mathrm{V}_{\mathrm{r}},\mathrm{V}_{\mathrm{t}})}&		\mathbf{H}_{m,n}^{(\mathrm{V}_{\mathrm{r}},\mathrm{H}_{\mathrm{t}})}\\
	\mathbf{H}_{m,n}^{(\mathrm{H}_{\mathrm{r}},\mathrm{V}_{\mathrm{t}})}&		\mathbf{H}_{m,n}^{(\mathrm{H}_{\mathrm{r}},\mathrm{H}_{\mathrm{t}})}\\
\end{matrix} \right] \in \mathbb{C} ^{2PQ\times 2PQ},
\end{equation}
where $\mathbf{H}_{m,n}^{(\mathrm{V}_{\mathrm{r}},\mathrm{V}_{\mathrm{t}})}\in \mathbb{C} ^{PQ\times PQ}$ denotes the sub-channel matrix between the vertical (V)-polarized transmit antennas and V-polarized receive antennas (for monostatic sensing scenario, the transmit antennas are also served as the receive antennas), and likewise for the other three blocks in (\ref{E_Dual_Polarized_Channel}).
For notational simplicity, let $\delta \in \left\{ \mathrm{V}_{\mathrm{r}},\mathrm{H}_r \right\} $
and $\eta \in \left\{ \mathrm{V}_t,\mathrm{H}_t \right\} $. Then, similar to the channel modeling in Section \ref{subsec_channel_modeling}, the $\left( \delta ,\eta \right)$
-th sub-channel matrix is modeled as
\begin{equation}
	\begin{aligned}
		&\ \mathbf{H}_{m,n}^{\delta ,\eta}\\
		=&\sum_{k=1}^K{\beta _{k}^{(\delta ,\eta )}}\mathbf{a}\left( \vartheta _k,\psi _k \right) \mathbf{a}\left( \vartheta _k,\psi _k \right) ^H\cdot e^{-j2\pi m\Delta f\tau _k}\cdot e^{j2\pi f_{k}^{d}nT_s},
	\end{aligned}
\end{equation}
where
\begin{equation}
	\beta _{k}^{(\delta ,\eta )}=\alpha _k\gamma _{k}^{(\delta ,\eta )},
\end{equation}
and $\gamma _{k}^{(\delta ,\eta )}$ denotes the polarization factor\cite{8481590}. In order to formulate the received signal into a forth-order tensor, we set the transmit precoding and the receive combining matrices for both V-polarized and H-polarized channels to be same. In this way, the equivalent transmit precoding and receive combining matrices for dual-polarization are respectively given by 
\begin{subequations}
	\begin{align}
		\tilde{\mathbf{F}}_{RX}=&\mathbf{I}_2\otimes \mathbf{F}_{RX}\in \mathbb{C} ^{2PQ\times 2R},\\
		\tilde{\mathbf{F}}_{TX}=&\mathbf{I}_2\otimes \mathbf{F}_{TX}\in \mathbb{C} ^{2PQ\times 2R}.
	\end{align}
\end{subequations}
To avoid the interference between dual-polarization, we assume that both dual-polarized channels share the same transmitted data. Then, the received signal vector of each polarization can be expressed as
\begin{equation}
	\mathbf{y}_{m,n}^{\delta}={\mathbf{F}}_{RX}^{H}\mathbf{H}_{m,n}^{\delta}\tilde{\mathbf{F}}_{TX}\tilde{\mathbf{e}}\cdot s_{m,n}+{\mathbf{F}}_{RX}^{H}\mathbf{n}_{m,n}^{\delta}\in \mathbb{C} ^{R\times 1},
\end{equation}
where $\mathbf{H}_{m,n}^{\delta}=\left[ \mathbf{H}_{m,n}^{\left( \delta ,\mathrm{V}_t \right)},\mathbf{H}_{m,n}^{\left( \delta ,\mathrm{H}_{\mathrm{t}} \right)} \right], \delta\in\{ \textrm{V}_r, \textrm{H}_r \}$, and $\tilde{\mathbf{e}}=[1,\dots,1]^T\in\mathbb{R}^{2R\times 1}$ denotes an all-one vector. Similarly, we still multiply the received signal vector by the conjugate of the transmitted data to eliminate its impacts, i.e.,
\begin{equation}
	\tilde{\mathbf{y}}_{m,n}^{\delta}=s_{m,n}^{*}\mathbf{y}_{m,n}^{\delta}={\mathbf{F}}_{RX}\mathbf{H}_{m,n}\tilde{\mathbf{F}}_{TX}\tilde{\mathbf{e}}+\tilde{\mathbf{n}}_{m,n}^{\delta}\in \mathbb{C} ^{R\times 1}.
\end{equation}
Then, stacking the received signal $\tilde{\mathbf{y}}_{m,n}^{\delta}$ among dual polarization, $N$ OFDM symbols and $M$ subcarriers into the following forth-order tensor as 
\begin{equation}
	\boldsymbol{\mathcal{Z}} =\sum_{k=1}^K{\alpha _k\mathbf{b}\left( \vartheta _k,\psi _k \right) \circ}\boldsymbol{\eta }_k \circ\mathbf{g}\left( \tau _k \right) \circ \mathbf{o}\left( f_{k}^{d} \right) +\boldsymbol{\mathcal{N}}, 
\end{equation}
where $\boldsymbol{\eta }_k=\left[ \gamma _{k}^{(\mathrm{V}_{\mathrm{r}},\mathrm{V}_{\mathrm{t}})}+\gamma _{k}^{(\mathrm{V}_{\mathrm{r}},\mathrm{H}_{\mathrm{t}})},\gamma _{k}^{(\mathrm{H}_{\mathrm{r}},\mathrm{V}_{\mathrm{t}})}+\gamma _{k}^{(\mathrm{H}_{\mathrm{r}},\mathrm{H}_{\mathrm{t}})} \right] ^T$, and the corresponding factor matrix introduced by the dual-polarization is given by
\begin{equation}
	\mathbf{B}=\left[ \boldsymbol{\eta} _1,\dots ,\boldsymbol{\eta} _K \right] \in \mathbb{C} ^{2\times K}.
\end{equation}
The mode-1 unfolding of $\boldsymbol{\mathcal{Z}}$ is given by
\begin{equation}
	\mathbf{Z}_{\left( 1 \right)}^{T}=\left( \mathbf{A}^{\left( 3 \right)}\odot \mathbf{A}^{\left( 2 \right)}\odot \mathbf{B} \right) \left( \mathbf{A}^{\left( 1 \right)} \right) ^T\in \mathbb{C} ^{2NM\times R}.
\end{equation}
Similarly, by defining the combining factor matrix as
\begin{equation}
	\mathbf{E}\triangleq \mathbf{A}^{\left( 2 \right)}\odot \mathbf{B},
\end{equation}
we can perform the tensor decomposition procedures proposed in Section \ref{sec_tensor_decomposition} to recover the factor matrices, i.e, ${\mathbf{A}}^{(1)}$, ${\mathbf{A}}^{(3)}$ and ${\mathbf{E}}$. The decoupling of ${\mathbf{A}}^{(2)}$ and ${\mathbf{B}}$ are as follows. Let $\hat{\mathbf{E}}_k=\mathrm{unvec}_{2\times N}\left( \hat{\mathbf{e}}_k \right) 
$, where $ \hat{\mathbf{e}}_k$ denotes the $k$-th column of $\hat{\mathbf{E}}$. Then, each column of $\mathbf{A}^{(2)}$ and $\mathbf{B}$ can be estimated by addressing the following problem as \cite{wang2024}
\begin{equation}
	\left\{ \hat{\mathbf{o}}_k,\hat{\boldsymbol{\eta}}_k \right\} =\mathrm{arg}\min_{{\mathbf{o}}_k,{\boldsymbol{\eta}}_k} \left\| \hat{\mathbf{E}}_k-{\boldsymbol{\eta}}_k{\mathbf{o}}_{k}^{T} \right\| _{F}^{2},
\end{equation}
which can be solved via performing the SVD of $\hat{\mathbf{E}}_k$. Specifically, according to Eckart–Young–Mirsky theorem\cite{golub1987generalization}, we have $
\hat{\boldsymbol{{\eta}}}_k=\lambda _{k,1}{\mathbf{u}}_{k,1}$ and $
\hat{\mathbf{o}}_k={\mathbf{v}}_{k,1}^{*}$, where $\lambda_{k,1}$,  $\mathbf{u}_{k,1}$ and ${\mathbf{v}}_{k,1}$ denote the maximum singular value of $\hat{\mathbf{E}}_k$, the corresponding left singular vector, and the corresponding right singular vector, respectively. Subsequently, we can estimate the multiple parameters from the recovered matrices as discussed in Section \ref{Subsec_parameters_est}. 

\section{Numerical Simulations}\label{sec_simulation}
In this section, simulation results are presented to evaluate the performance of the proposed schemes. We first provide the simulation parameter settings. Then, we compare the performance of the proposed monostatic parameter estimation scheme with that of conventional techniques. Subsequently, we demonstrate the performance of the cooperative position and velocity estimation. \textcolor{black}{Finally, we evaluate the generality of the proposed scheme in a degraded scenario and compare its performance with the state-of-the-art scheme.}

\subsection{Simulation Settings}
\begin{figure}[t]
	\centering
	\includegraphics[width=1.5in]{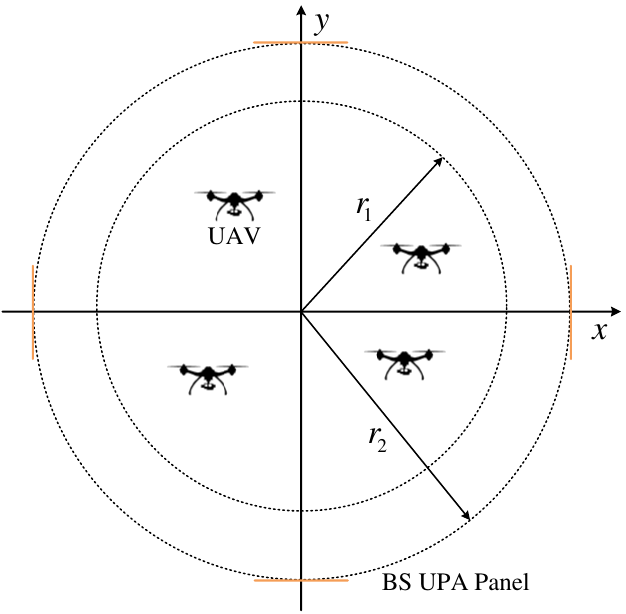}
	\caption{The top view of simulation layout.}
	\label{fig_simu_layout}
\end{figure}
Unless stated otherwise, the simulation parameters are set as follows: Each BS is equipped with a half-wavelength UPA with $P=\textrm{16}$ and $Q=\textrm{24}$ antennas located in horizontal and vertical direction, respectively. The number of RF chains is given by $R=\textrm{64}$. \textcolor{black}{The non-zero elements of the precoding matrix is designed with the guidelines of the prior information provided in Section \ref{sec_pre_steps}. The non-zeros elements of the combining matrix are chosen uniformly from a normalized unit circle to guarantee the uniqueness of tensor decomposition \cite{7914672} and the unambiguity of AoA estimation in the HBF structure.} In order to avoid the interference between the BSs, we assume that each BS is allocated a non-overlapping 20 MHz bandwidth with 612 subcarriers. We set the central frequency of 4.9 GHz and the SCS of $\Delta f =\textrm{30}$ KHz. The number of OFDM symbols utilized for sensing is set to $N=\textrm{7}$. According to 5G new radio (NR) standard, the total period of OFDM symbol (including the cyclic prefix) is given by $T_s = \textrm{35.677} \ \mathrm{\mu s}$\cite{3GPP13810101}. The velocity of each UAV is uniformly distributed in $[V_{\min},V_{\max}]$, where $V_{\min}=\textrm{5}\ \textrm{km/h}$ and $V_{\max}=\textrm{100}\ \textrm{km/h}$ denote the minimum and the maximum velocity of UAV, respectively. The RCS of each UAV is set to $\sigma=\textrm{0.01}\ \textrm{m}^2$. The path loss in dB is given by\cite{ITU}
\begin{equation}
	\mathrm{PL}=\textrm{103.4}+\textrm{20}\lg \left( f/\mathrm{MHz} \right) +\textrm{40}\lg \left( d/\mathrm{km} \right) -\textrm{10}\lg \left( \sigma /\mathrm{m}^2 \right),
\end{equation}
where $f$ denotes the central frequency, $d$ denotes the range between the UAV and the BS. The noise power density is set to -174 dBm/Hz. For the simulation layout, as depicted in Fig. \ref{fig_simu_layout}, \textcolor{black}{there are $J=4$ BSs uniformly situated on a circle with the radius of $r_2=\textrm{450}$ m}, with their UPA panels point towards the center of the circle, and the height of each BS is set to 30 m. The transmit power budget of each BS is set to 58 dBm. In addition, \textcolor{black}{it is assumed that there are $K=\textrm{4}$ UAVs uniformly distributed with their projections on the $x$-$y$ plane within a circle with the radius of $r_1=\textrm{400}$ m}, and the heights of UAVs are uniformly distributed in $[h_{\min},h_{\max}]$, where $h_{\min}=\textrm{35}$ m and $h_{\max}=\textrm{300}$ m denote the lowest and highest flight heights of UAVs, respectively. To evaluate the performance of parameter estimation, the following root mean square error (RMSE) is adopted\cite{10403776}, i.e.,
\begin{equation}
	\mathrm{RMSE}\left( \mathbf{x} \right) =\sqrt{\frac{1}{K}\sum_{k=1}^K{\left\| \hat{\mathbf{x}}_k-\mathbf{x}_k \right\| _{2}^{2}}}, 
\end{equation}
where $\hat{\mathbf{x}}_k$ and $\mathbf{x}_k$ respectively denote the estimated and the true value of parameters, including AoAs, range, radial velocity, position and true velocity. The simulation results are obtained by averaging over \textcolor{black}{95\%} of more than 1000 independent realizations to ignore the effect of outliers\cite{10403776}. \textcolor{black}{Additionally, we consider a conventional approach relying on MUSIC and FFT algorithms as a benchmark (denoted as ``Benchmark1"). Specifically, in this scheme, the UAVs' AoAs are first derived by the well-known 2-D MUSIC algorithm\cite{17564}. Then, the 2-D FFT operation is performed to estimate the ranges and radial velocities of UAVs\cite{han2024cellular}. To ensure the automatic pairing of multi-dimensional parameters, this benchmark also operates on the formulation of the mode-1 unfolding of tensor, i.e., (\ref{E_Mode1_unfolding}).
}

\begin{figure}[t]
	\centering
	\subfigure[AoA]{\includegraphics[width=0.22\textwidth]{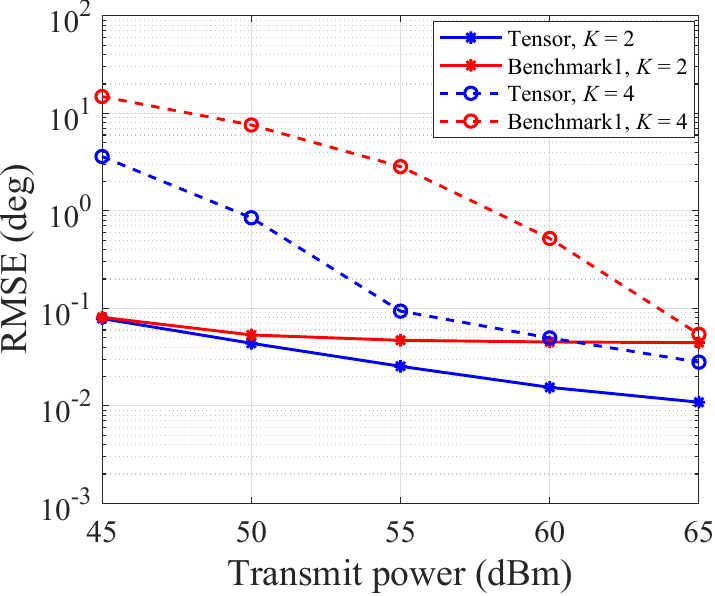}\label{fig_angle_RMSE}}
	\ 
	\subfigure[Range]{\includegraphics[width=0.22\textwidth]{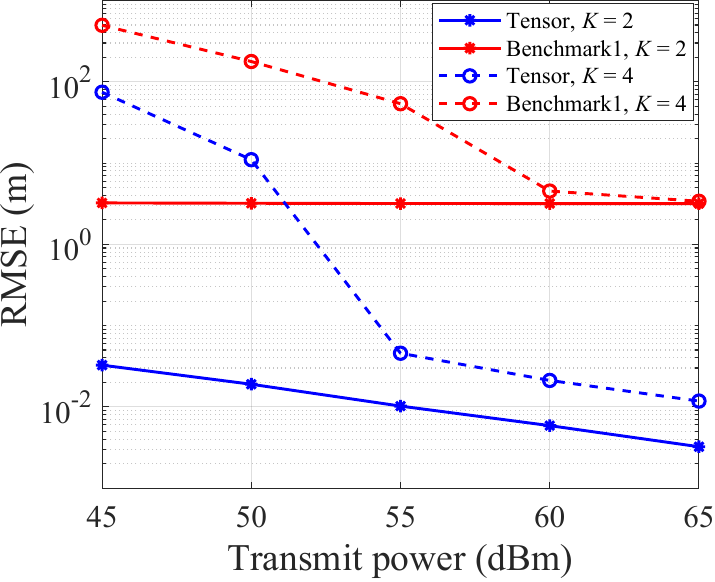}\label{fig_range_RMSE}}
	\
	\subfigure[Radial velocity]{\includegraphics[width=0.22\textwidth]{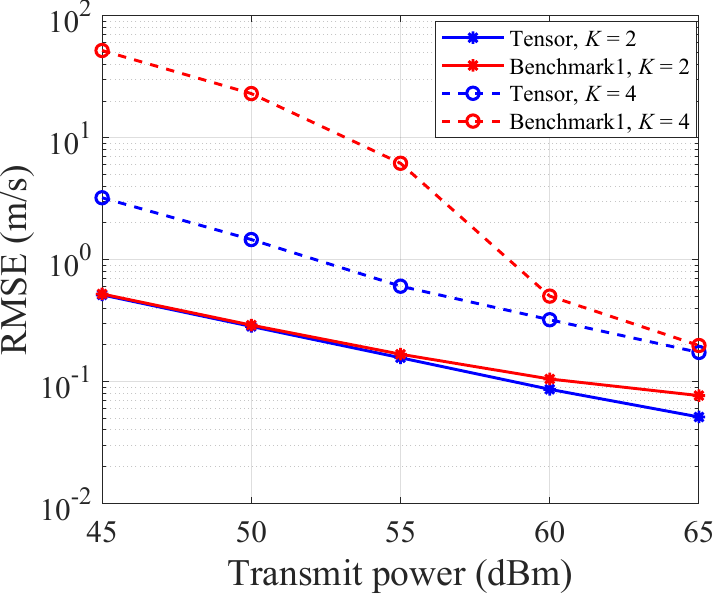}\label{fig_radial_velocity_RMSE}}
	\
	\subfigure[Position]{\includegraphics[width=0.22\textwidth]{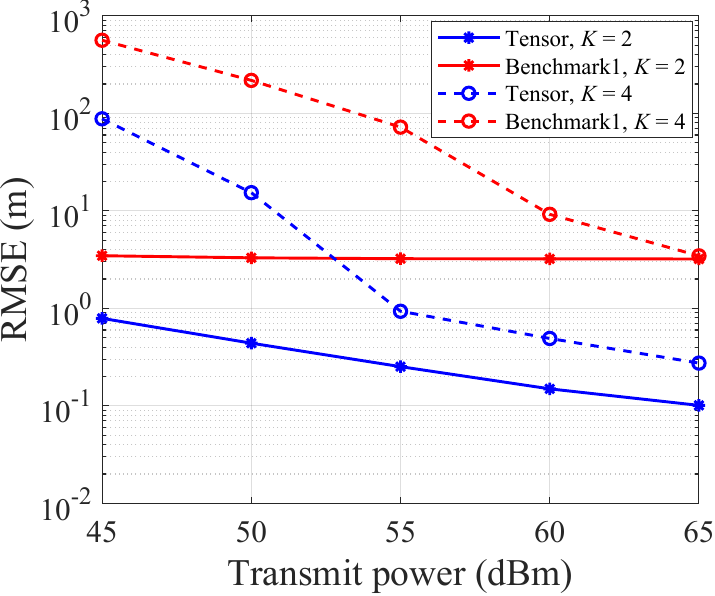}\label{fig_position_RMSE_for_single_BS}}
	\caption{Estimation RMSE vs. the transmit power budget.}	
\end{figure}\label{fig_mono_estimation_RMSE}
\subsection{Monostatic Parameter Estimation}
According to the above discussions, since each BS performs the same parameter estimation algorithm before data fusion, we first set $J=1$  to evaluate the performance of the proposed monostatic parameter estimation scheme. 

Figs. \ref{fig_angle_RMSE}-\ref{fig_radial_velocity_RMSE} respectively illustrate the RMSE of AoA, range and radial velocity estimation versus the transmit power budget. From Figs. \ref{fig_angle_RMSE}-\ref{fig_radial_velocity_RMSE}, we observe that as the transmit power increases, both two schemes achieve lower estimation RMSE. In addition, due to limitations in antenna array size and sensing resource allocation, we find that the estimation RMSE of Benchmark1 is prone to encountering a performance bottleneck. However, the estimation RMSE of the proposed scheme shows a significant decline with the increase of transmit power, which validates its effectiveness. Correspondingly, Fig. \ref{fig_position_RMSE_for_single_BS} illustrates that the proposed tensor decomposition scheme achieves higher positioning accuracy than Benchmark1, since the former achieves enhanced AoA and range estimation performance.

\begin{figure}[t]
	\centering
	\includegraphics[width=3.0in]{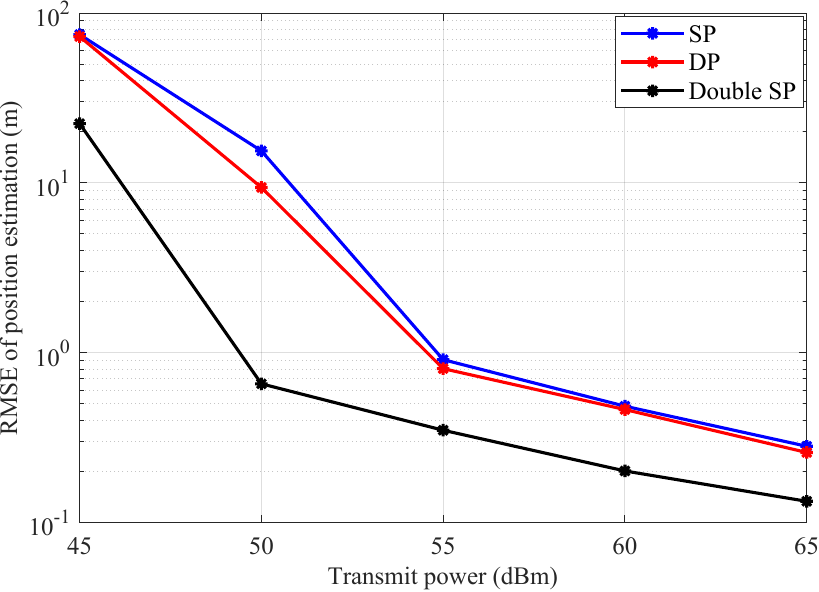}
	\caption{Position estimation RMSE vs. the transmit power budget in the dual-polarized system.}
	\label{fig_DP_position_RMSE}
\end{figure}
In addition, we also provide the position estimation results in the dual-polarized system (denoted as ``DP") and compare with the single-polarized configuration (denoted as ``SP") in Fig. \ref{fig_DP_position_RMSE} with the number of UAVs of $K = \textrm{4}$. The parameters of DP configuration are set as follows. For sake of clarity, we rewrite the polarization factor as
\begin{equation}
	\gamma _{k}^{(\delta ,\eta )}=\sqrt{r_{k}^{(\delta ,\eta )}}e^{j\phi _{k}^{(\delta ,\eta )}},
\end{equation}
where $r_{k}^{(\delta ,\eta )}$ denotes the random variable representing the power ratio of waves from $\eta =\left\{ \mathrm{V}_t,\mathrm{H}_t \right\}$ transmit antennas to $\delta \in \left\{ \mathrm{V}_{\mathrm{r}},\mathrm{H}_r \right\}$ receive antennas, $\phi _{k}^{(\delta ,\eta)}$ denotes the additional phase. According to the 3rd generation partnership project (3GPP) technical report (TR) \cite{3GPP_DP}, we let $r_{k}^{(\mathrm{V}_r,\mathrm{V}_{\mathrm{t}})}=r_{k}^{(\mathrm{H}_r,\mathrm{H}_{\mathrm{t}})}=1$ and $r_{k}^{(\mathrm{V}_r,\mathrm{H}_{\mathrm{t}})},r_{k}^{(\mathrm{H}_r,\mathrm{V}_{\mathrm{t}})}=1/\xi$, where $\xi$ denotes the cross-polarization discrimination (XPD) obeying the Log-Normal distribution, i.e., $\xi\sim \mathcal{N}\left( \mu ,\sigma ^2 \right)$ dB with the expectation of $\mu =\textrm{8}$ and the standard deviation of $\sigma = \textrm{4}$. The phases are assumed to be random variables obeying the uniform distribution, i.e., $\phi _{k}^{(\delta ,\eta )}\sim \mathcal{U} \left( 0,2\pi \right)$. To guarantee the fairness, we also incorporate the double SP configuration (denoted as ``Double SP"). Specifically, we set the number of RF chains as $\tilde{R}=2R=\textrm{128}$, the number of antennas located in horizontal direction as $\tilde{P}=2P=\textrm{32}$ in this configuration. Fig. \ref{fig_DP_position_RMSE} illustrates that DP configuration achieves higher positioning accuracy than the SP configuration, since the DP configuration provides more abundant amount of samples for parameter estimation. However, the double SP configuration achieves even higher positioning accuracy than the DP configuration due to enhanced AoA estimation achieved by more sufficient array manifold information and reduced cumulative error. Nevertheless, the employment of dual polarization is still promising, since it reduces the antenna deployment area compared to the double SP configuration.

\begin{figure}[t]
	\centering
	\includegraphics[width=3.0in]{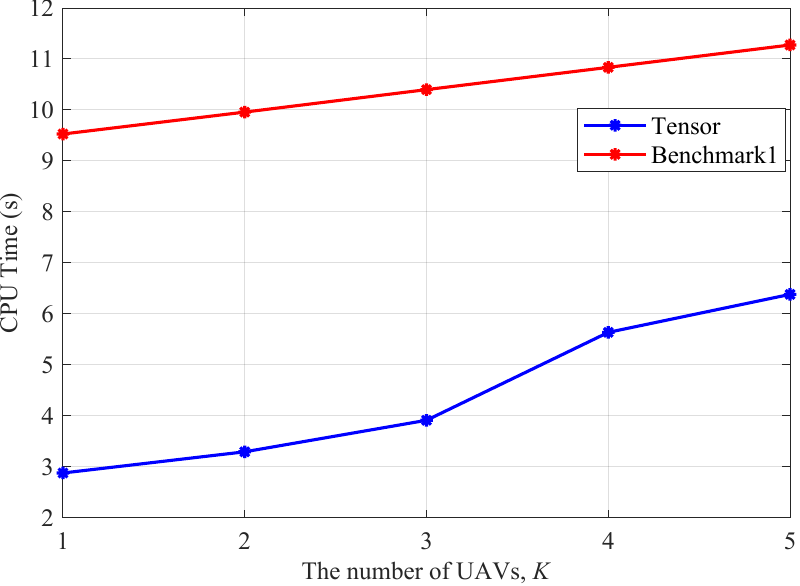}
	\caption{\textcolor{black}{The CPU time of the monostatic parameter estimation.}}
	\label{fig_CPU_Time}
\end{figure}

Fig. \ref{fig_CPU_Time} depicts the central processing unit (CPU) time of the monostatic parameter estimation versus the number of UAVs. It can be observed that the proposed tensor decomposition-based scheme is more efficient than Benchmark1. The reason lies in the fact that the MUSIC algorithm requires the 2-D search on $(\theta,\phi)$ for AoA estimation, whereas for the tensor decomposition scheme, the proposed GRQ-based AoA estimation algorithm can effectively reduce the computational complexity. In addition, we also find that with increased $K$, the CPU time for both two schemes increases as more calculations are required to estimate more parameters.

\begin{figure}[t]
	\centering
	\includegraphics[width=3.0in]{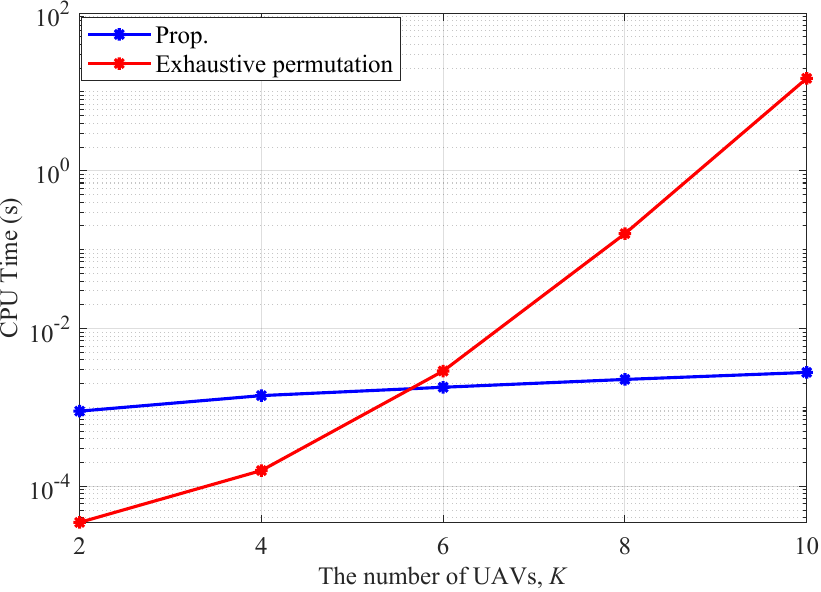}
	\caption{\textcolor{black}{The CPU time of the multi-BS data association.}}
	\label{fig_coop_data_asso_cpu_time}
\end{figure}
\begin{figure}[t]
	\centering
	\includegraphics[width=3.0in]{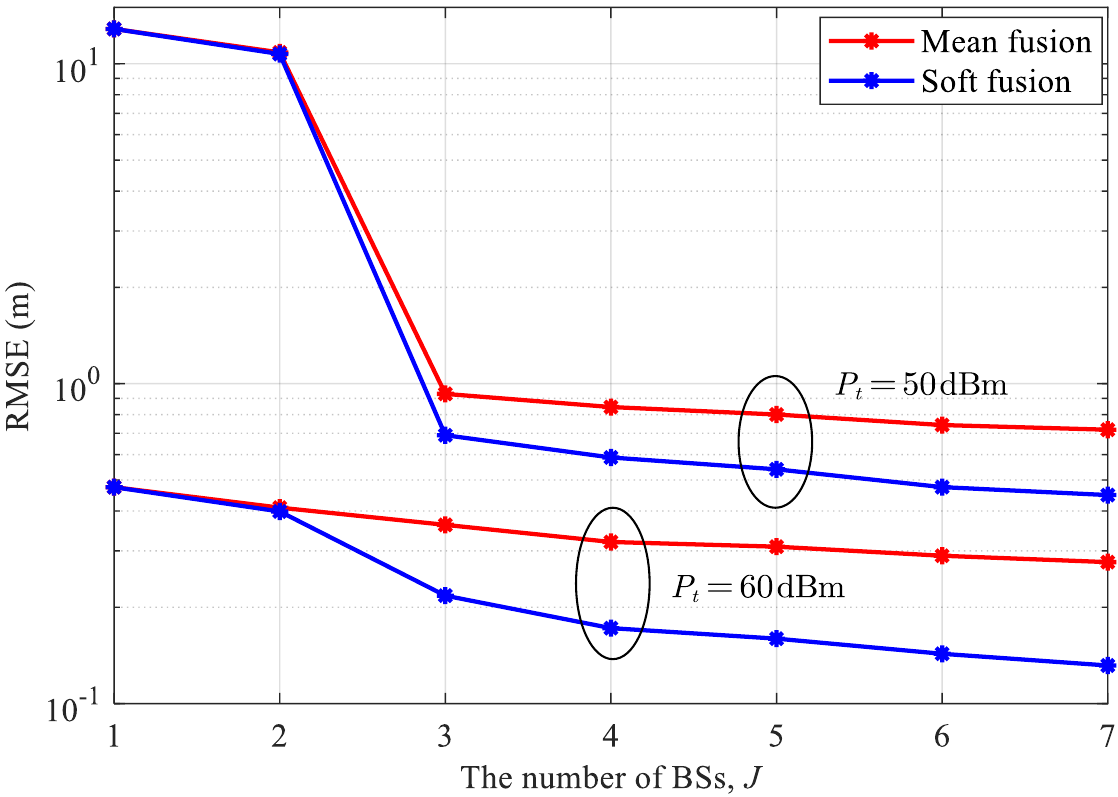}
	\caption{\textcolor{black}{Cooperative position estimation RMSE vs. the number of BSs.}}
	\label{fig_coop_position_RMSE}
\end{figure}

\subsection{Cooperative Position and Velocity Estimation}\label{subsec_Coop_simu}
\textcolor{black}{Fig. \ref{fig_coop_data_asso_cpu_time} illustrates the efficiency of the proposed multi-BS data association algorithm. It can be observed that with a small number of UAVs, the CPU time of the proposed algorithm exceeds that of the exhaustive permutation scheme. This is because, in addition to deriving the MST of the updated graph $\tilde{G}$, the proposed algorithm needs additional procedures to calculate the distances between the positioning results of each BS and those of other BSs to recognize and remove the false detection results to enhance the cooperative sensing scheme. Moreover, as the number of UAVs increases, the computational complexity of the exhaustive permutation method rises sharply, while the CPU time of the proposed algorithm increases slowly, which allows the algorithm to perform well even when there are many UAVs.}

Fig. \ref{fig_coop_position_RMSE} depicts the cooperative position estimation RMSE versus the number of BSs with the weighting intensity factor of $\beta_1 =\textrm{0.5}$. As previously outlined, our strategy for selecting the lattice point from the Pareto set is contingent upon the system configuration and the precision of AoA and range estimation. Specifically, for the tensor decomposition scheme, noting that we assign 612 subcarriers but only 384 antennas (with HBF structure) to each BS. In addition, the AoA estimation will also be degraded by the cumulative error effect, which makes the estimation of range more precise than AoA. Thus, we choose the lattice point from the Pareto set according to (\ref{E_range_dominant}). From Fig. \ref{fig_coop_position_RMSE}, we make the following observations: First, we observe that the increase of BSs leads to a continuous improvement in cooperative positioning accuracy, which is significantly higher than the monostatic positioning results (i.e., $K = 1$). This is attributed to the fact that more BSs provide richer UAVs' parameter estimations, i.e., AoAs and range estimations for the data fusion. Furthermore, it also can be observed that the proposed Pareto optimality scheme further reduces the positioning error compared to the mean fusion scheme. This is because it takes the AoA and range estimations into account with a finer granularity, rather than merely fuse the final positioning results. In addition, it also allocates distinct weights to the BSs according to the ranges between them and the UAVs, making the more accurate estimations more dominant in the fusion process.

\begin{figure}[t]
	\centering
	\includegraphics[width=3.0in]{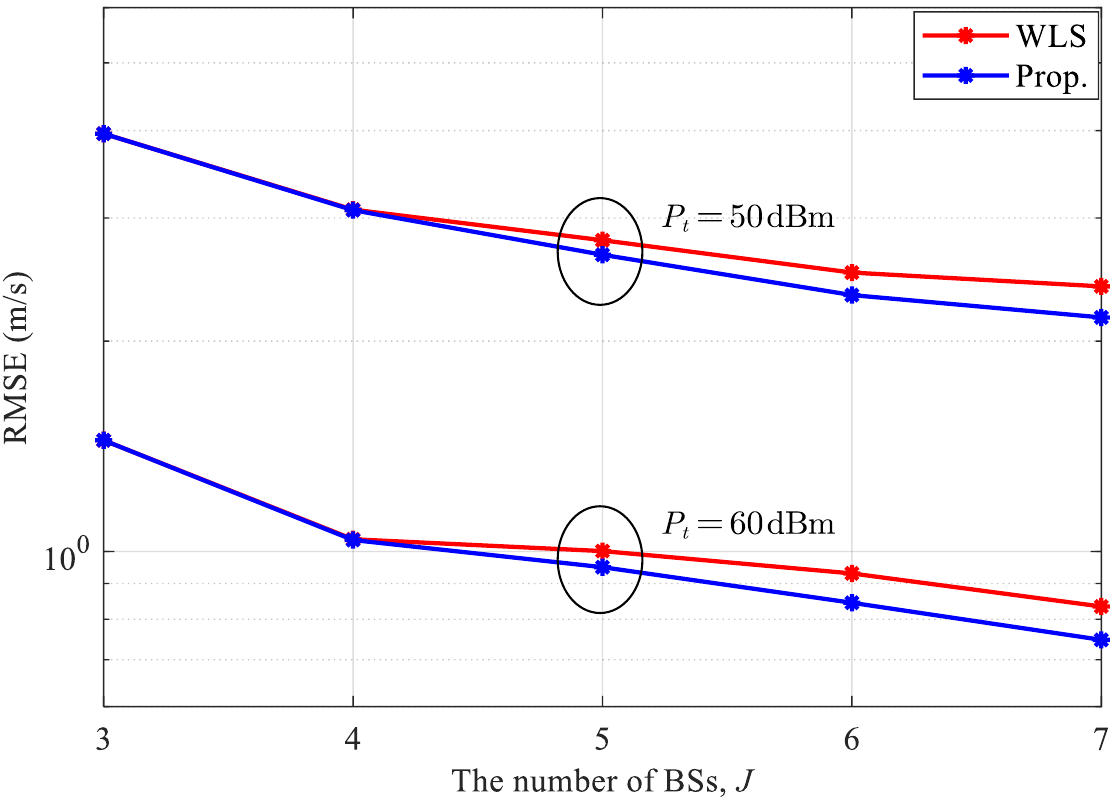}
	\caption{\textcolor{black}{Cooperative true velocity estimation RMSE vs. the number of BSs.}}
	\label{fig_coop_Velocity_RMSE}
\end{figure}
Fig. \ref{fig_coop_Velocity_RMSE} shows the cooperative velocity estimation RMSE versus the number of BSs with the weighting intensity factor of $\beta_2 = \textrm{0.5}$. Similarly, it can be observed again from Fig. \ref{fig_coop_Velocity_RMSE} that the velocity estimation error decreases as the number of BSs increases. This is because more BSs provide more information about the radial velocities and AoAs. Furthermore, the proposed residual weighting-based scheme further facilitates the velocity estimation compared to the WLS method due to the suppression of estimation errors. 

\subsection{\textcolor{black}{Performance Evaluation in the Degraded Scenario}}
\textcolor{black}{To further evaluate the versatility and generality of the proposed sensing scheme, we degrade it to the single-antenna  single-target scenario and compare its performance with the state-of-the-art cooperative sensing scheme proposed in \cite{10226276} (denoted as ``Benchmark2"). In the degraded scenario, the following adjustments are performed:}
\begin{enumerate}
	\item \textcolor{black}{The parameter estimation problem formulated in (\ref{E_TALS}) degenerates into a second-order tensor decomposition model, while the Vandermonde property of the factor matrix can still be utilized to recover the factor matrices and estimate the UAV's range and radial velocity via the similar procedures provided in Section \ref{sec_tensor_decomposition}.}

	\item \textcolor{black}{Without AoA estimation, the rough estimation of the UAV's position is derived by the LS method\cite{10226276}.}
	
	\item \textcolor{black}{Without AoA estimation, the determination of the final position estimation from the lattice points union should only take the minimization of the range loss function into consideration, i.e., (\ref{E_range_dominant}).} 
\end{enumerate}

\begin{figure}[t]
	\centering
	\subfigure[\textcolor{black}{Range}]{\includegraphics[width=0.22\textwidth]{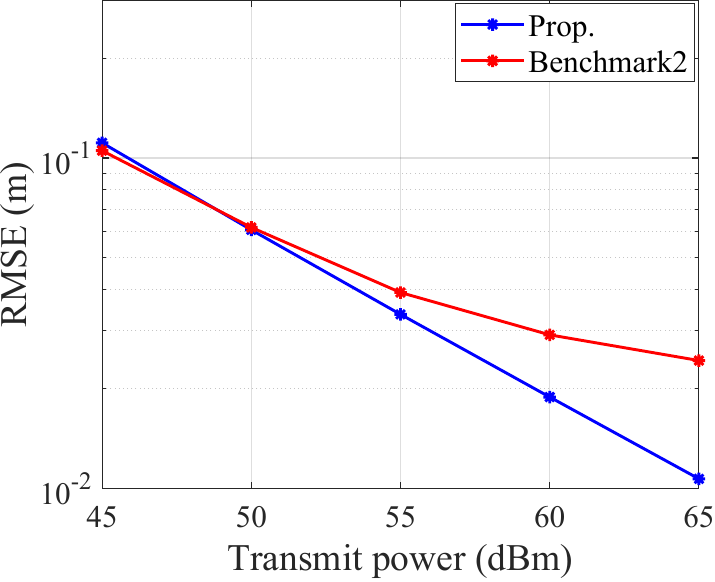}\label{fig_SA_Range_RMSE}}
	\
	\subfigure[\textcolor{black}{Radial velocity}]{\includegraphics[width=0.22\textwidth]{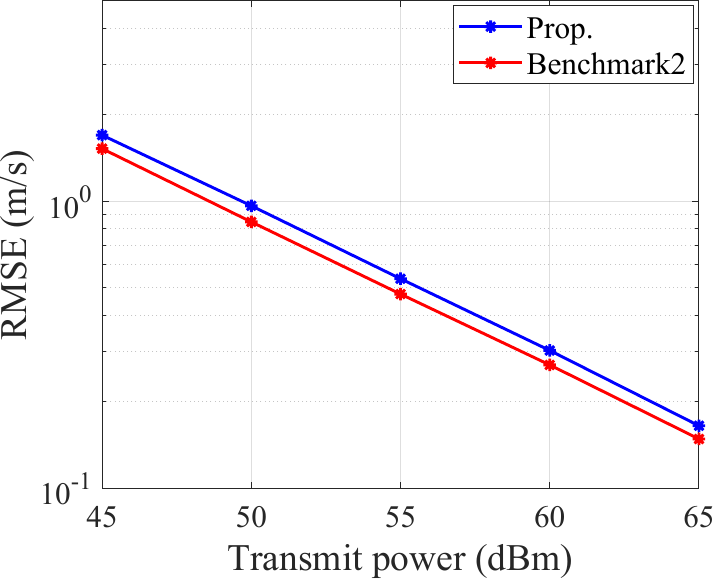}\label{fig_SA_Radial_Velocity_RMSE}}
	\
	\subfigure[\textcolor{black}{Position}]{\includegraphics[width=0.22\textwidth]{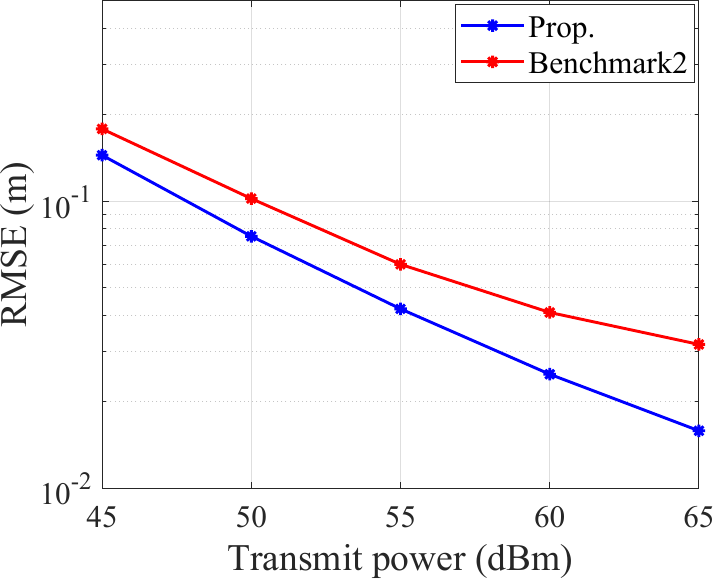}\label{fig_Coop_Position_RMSE}}
	\
	\subfigure[\textcolor{black}{True Velocity}]{\includegraphics[width=0.22\textwidth]{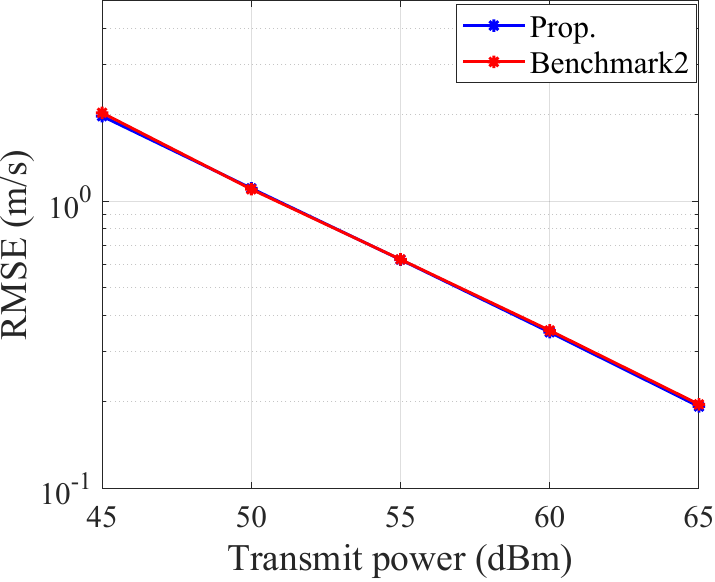}\label{fig_Coop_Velocity_RMSE}}
	\caption{\textcolor{black}{Estimation RMSE vs. the transmit power budget in the degraded scenario.}}	\label{fig_SA_mono_estimation_RMSE}
\end{figure}

\textcolor{black}{Fig. \ref{fig_SA_mono_estimation_RMSE} illustrates the sensing performance of the proposed scheme and of Benchmark2. It should be noted that in the considered single-antenna setup, due to the severe path loss and the limited antenna gain, it is hard to achieve sufficient received SNR to meet the UAV detection and parameter estimation requirements, especially when the UAV is far from the BSs. Thus, in this simulation, we adjust the simulation parameters as $r_1=100 \ \textrm{m}, r_2 = 125\  \textrm{m}$. In addition, for ease of algorithm implementation, we assume that the UAV and the BSs are situated on the same horizontal plane, and its velocity component in the $z$-axis is set to zero. From Fig. \ref{fig_SA_mono_estimation_RMSE}, we find that the proposed scheme demonstrates significant performance gains in range and position estimation compared to Benchmark2, due to its off-grid approach to directly derive the range estimation relying on the Vandermonde structure of the factor matrix. In terms of radial velocity estimation, the proposed scheme slightly performs worse than Benchmark2, attributed to the cumulative effect of errors, since the radial velocity estimation is affected by the previous range estimation. However, due to the superior position estimation, the proposed solution still achieves comparable performance in true velocity estimation. Additionally, different from Benchmark2's symbol-level multi-BS fusion approach, the proposed data fusion scheme requires each BS to upload fewer sensing parameters to the cloud, thereby reducing the transmission overhead and synchronization requirements.}

\section{Conclusion}\label{sec_conclusion}
In this paper, we proposed a comprehensive cooperative ISAC scheme for the low-altitude sensing scenario that includes the monostatic parameter estimation, multi-BS data association, and multi-BS cooperative sensing. Specifically, we first provided preliminary steps for the sensing scheme and formulated the monostatic parameter estimation problem via using a tensor decomposition model to estimate the UAVs' parameters. Then, a false removing MST-based data association method was developed to accurately match the multiple BSs' estimations to the same UAV. Subsequently, we proposed a Pareto optimality method and a residual weighting scheme to improve the position and velocity estimation, respectively. Additionally, we also extended our approach to the dual-polarized system. Simulation results demonstrated the superiority of proposed schemes in terms of generality and estimation accuracy to the conventional techniques. 


\bibliographystyle{IEEEtran}
\bibliography{Reference}

\begin{thebibliography}{10}
\providecommand{\url}[1]{#1}
\csname url@samestyle\endcsname
\providecommand{\newblock}{\relax}
\providecommand{\bibinfo}[2]{#2}
\providecommand{\BIBentrySTDinterwordspacing}{\spaceskip=0pt\relax}
\providecommand{\BIBentryALTinterwordstretchfactor}{4}
\providecommand{\BIBentryALTinterwordspacing}{\spaceskip=\fontdimen2\font plus
\BIBentryALTinterwordstretchfactor\fontdimen3\font minus
  \fontdimen4\font\relax}
\providecommand{\BIBforeignlanguage}[2]{{%
\expandafter\ifx\csname l@#1\endcsname\relax
\typeout{** WARNING: IEEEtran.bst: No hyphenation pattern has been}%
\typeout{** loaded for the language `#1'. Using the pattern for}%
\typeout{** the default language instead.}%
\else
\language=\csname l@#1\endcsname
\fi
#2}}
\providecommand{\BIBdecl}{\relax}
\BIBdecl

\bibitem{9456851}
Q.~Wu, J.~Xu, Y.~Zeng, D.~W.~K. Ng, N.~Al-Dhahir, R.~Schober, and A.~L.
  Swindlehurst, ``A comprehensive overview on {5G}-and-beyond networks with
  {UAVs}: From communications to sensing and intelligence,'' \emph{IEEE J. Sel.
  Areas Commun.}, vol.~39, no.~10, pp. 2912--2945, Oct. 2021.

\bibitem{9737357}
F.~Liu, Y.~Cui, C.~Masouros, J.~Xu, T.~X. Han, Y.~C. Eldar, and S.~Buzzi,
  ``Integrated sensing and communications: Toward dual-functional wireless
  networks for {6G} and beyond,'' \emph{IEEE J. Sel. Areas Commun.}, vol.~40,
  no.~6, pp. 1728--1767, Jun. 2022.

\bibitem{9906898}
S.~Lu, F.~Liu, and L.~Hanzo, ``The degrees-of-freedom in monostatic {ISAC}
  channels: {NLoS} exploitation vs. reduction,'' \emph{IEEE Trans. Veh.
  Technol.}, vol.~72, no.~2, pp. 2643--2648, Feb. 2023.

\bibitem{9724260}
L.~Pucci, E.~Paolini, and A.~Giorgetti, ``System-level analysis of joint
  sensing and communication based on {5G} new radio,'' \emph{IEEE J. Sel. Areas
  Commun.}, vol.~40, no.~7, pp. 2043--2055, Jul. 2022.

\bibitem{8318564}
C.~Cui, J.~Xu, R.~Gui, W.-Q. Wang, and W.~Wu, ``Search-free {DOD}, {DOA} and
  range estimation for bistatic {FDA-MIMO} radar,'' \emph{IEEE Access}, vol.~6,
  pp. 15\,431--15\,445, Mar. 2018.

\bibitem{9860521}
L.~Leyva, D.~Castanheira, A.~Silva, and A.~Gameiro, ``Two-stage estimation
  algorithm based on interleaved {OFDM} for a cooperative bistatic {ISAC}
  scenario,'' in \emph{Proc. IEEE Veh. Technol. Conf. (VTC)}, Jun. 2022, pp.
  1--6.

\bibitem{han2024cellular}
Z.~Han, H.~Ding, L.~Han, L.~Ma, X.~Zhang, M.~Lou, Y.~Wang, J.~Jin, Q.~Wang,
  G.~Liu \emph{et~al.}, ``Cellular network based multistatic integrated sensing
  and communication systems,'' \emph{IET Commun.}, vol.~6, pp. 1--11, Jan.
  2024.

\bibitem{10464728}
Z.~Han, H.~Ding, X.~Zhang, Y.~Wang, M.~Lou, J.~Jin, Q.~Wang, and G.~Liu,
  ``Multistatic integrated sensing and communication system in cellular
  networks,'' in \emph{Proc. IEEE Globecom Workshops (GC Wkshps)}, Dec. 2023,
  pp. 123--128.

\bibitem{9916293}
G.~Li, S.~Wang, K.~Ye, M.~Wen, D.~W.~K. Ng, and M.~Di~Renzo, ``Multi-point
  integrated sensing and communication: Fusion model and functionality
  selection,'' \emph{IEEE Wireless Commun. Lett.}, vol.~11, no.~12, pp.
  2660--2664, Dec. 2022.

\bibitem{10032141}
P.~Gao, L.~Lian, and J.~Yu, ``Cooperative {ISAC} with direct localization and
  rate-splitting multiple access communication: A pareto optimization
  framework,'' \emph{IEEE J. Sel. Areas Commun.}, vol.~41, no.~5, pp.
  1496--1515, May. 2023.

\bibitem{10207026}
Y.~Cao and Q.-Y. Yu, ``Joint resource allocation for user-centric cell-free
  integrated sensing and communication systems,'' \emph{IEEE Commun. Lett.},
  vol.~27, no.~9, pp. 2338--2342, Sep. 2023.

\bibitem{9724258}
Q.~Shi, L.~Liu, S.~Zhang, and S.~Cui, ``Device-free sensing in {OFDM} cellular
  network,'' \emph{IEEE J. Sel. Areas Commun.}, vol.~40, no.~6, pp. 1838--1853,
  Jun. 2022.

\bibitem{zhang2024}
Z.~Zhang, H.~Ren, C.~Pan, S.~Hong, D.~Wang, J.~Wang, and X.~You, ``Target
  localization in cooperative {ISAC} systems: A scheme based on {5G} {NR}
  {OFDM} signals,'' \emph{IEEE Trans. Commun.}, pp. 1--1, 2024.

\bibitem{10226276}
Z.~Wei, R.~Xu, Z.~Feng, H.~Wu, N.~Zhang, W.~Jiang, and X.~Yang, ``Symbol-level
  integrated sensing and communication enabled multiple base stations
  cooperative sensing,'' \emph{IEEE Trans. Veh. Technol.}, vol.~73, no.~1, pp.
  724--738, Jan. 2024.

\bibitem{10615952}
X.~Lu, Z.~Wei, R.~Xu, L.~Wang, B.~Lu, and J.~Piao, ``Integrated sensing and
  communication enabled multiple base stations cooperative {UAV} detection,''
  in \emph{Proc. IEEE Int. Conf. on Commun. Workshops (ICC Workshops)}, Jun.
  2024, pp. 1882--1887.

\bibitem{10403776}
R.~Zhang, L.~Cheng, S.~Wang, Y.~Lou, Y.~Gao, W.~Wu, and D.~W.~K. Ng,
  ``Integrated sensing and communication with massive {MIMO}: A unified tensor
  approach for channel and target parameter estimation,'' \emph{IEEE Trans.
  Wireless Commun.}, vol.~23, no.~8, pp. 8571--8587, Aug. 2024.

\bibitem{7914672}
Z.~Zhou, J.~Fang, L.~Yang, H.~Li, Z.~Chen, and R.~S. Blum, ``Low-rank tensor
  decomposition-aided channel estimation for millimeter wave {MIMO-OFDM}
  systems,'' \emph{IEEE J. Sel. Areas Commun.}, vol.~35, no.~7, pp. 1524--1538,
  Jul. 2017.

\bibitem{9049103}
Y.~Lin, S.~Jin, M.~Matthaiou, and X.~You, ``Tensor-based channel estimation for
  millimeter wave {MIMO-OFDM} with dual-wideband effects,'' \emph{IEEE Trans.
  Commun.}, vol.~68, no.~7, pp. 4218--4232, Jul. 2020.

\bibitem{wang2024}
R.~Wang, H.~Ren, C.~Pan, G.~Zhou, and J.~Wang, ``Tensor decomposition-based
  time varying channel estimation for mmwave {MIMO-OFDM} systems,'' 2024,
  \emph{arXiv:2403.02942}. [Online]. Available:
  https://arxiv.org/abs/2403.02942.

\bibitem{5281762}
H.~Zhu and J.~Wang, ``Chunk-based resource allocation in {OFDMA} systems—part
  {I}: Chunk allocation,'' \emph{IEEE Trans. Commun.}, vol.~57, no.~9, pp.
  2734--2744, Sep. 2009.

\bibitem{6094142}
H.~Zhu and J.~Wang, ``Chunk-based resource allocation in {OFDMA} systems—part
  {II}: Joint chunk, power and bit allocation,'' \emph{IEEE Trans. Commun.},
  vol.~60, no.~2, pp. 499--509, Feb. 2012.

\bibitem{9746355}
Y.~Cui, X.~Jing, and J.~Mu, ``Integrated sensing and communications via {5G}
  {NR} waveform: Performance analysis,'' in \emph{Proc. IEEE Int. Conf. Acoust.
  Speech Signal Process. (ICASSP)}, May 2022, pp. 8747--8751.

\bibitem{9201513}
F.~Liu and C.~Masouros, ``A tutorial on joint radar and communication
  transmission for vehicular networks—part {I}: Background and
  fundamentals,'' \emph{IEEE Commun. Lett.}, vol.~25, no.~2, pp. 322--326, Feb.
  2021.

\bibitem{Sta_signal_process}
S.~M. Kay, \emph{Fundamentals of Statistical Signal Processing}.\hskip 1em plus
  0.5em minus 0.4em\relax Englewood Cliffs, NJ, USA: Prentice-Hall, 1998.

\bibitem{1164557}
M.~Wax and T.~Kailath, ``Detection of signals by information theoretic
  criteria,'' \emph{IEEE Trans. Acoust. Speech Signal Process.}, vol.~33,
  no.~2, pp. 387--392, Apr. 1985.

\bibitem{tensor}
\BIBentryALTinterwordspacing
T.~G. Kolda and B.~W. Bader, ``Tensor decompositions and applications,''
  \emph{SIAM Review}, vol.~51, no.~3, pp. 455--500, 2009. [Online]. Available:
  \url{https://doi.org/10.1137/07070111X.}
\BIBentrySTDinterwordspacing

\bibitem{zxd}
X.~Zhang, \emph{Matrix analysis and applications}.\hskip 1em plus 0.5em minus
  0.4em\relax Beijing, CHN: Tsinghua University Press, 2004.

\bibitem{6573422}
M.~Sørensen and L.~De~Lathauwer, ``Blind signal separation via tensor
  decomposition with vandermonde factor: Canonical polyadic decomposition,''
  \emph{IEEE Trans. Signal Process.}, vol.~61, no.~22, pp. 5507--5519, Nov.
  2013.

\bibitem{Kruskal}
\BIBentryALTinterwordspacing
J.~B. Kruskal, ``Three-way arrays: rank and uniqueness of trilinear
  decompositions, with application to arithmetic complexity and statistics,''
  \emph{Linear Algebra and its Applications}, vol.~18, no.~2, pp. 95--138,
  1977. [Online]. Available:
  \url{https://www.sciencedirect.com/science/article/pii/0024379577900696.}
\BIBentrySTDinterwordspacing

\bibitem{9583869}
X.~Zhang, F.~Wang, and H.~Li, ``An efficient method for cooperative
  multi-target localization in automotive radar,'' \emph{Signal Process.
  Lett.}, vol.~29, pp. 16--20, Oct. 2022.

\bibitem{6773228}
R.~C. Prim, ``Shortest connection networks and some generalizations,''
  \emph{Bell Syst. Tech. J.,}, vol.~36, no.~6, pp. 1389--1401, Nov. 1957.

\bibitem{kruskal1956shortest}
J.~B. Kruskal, ``On the shortest spanning subtree of a graph and the traveling
  salesman problem,'' in \emph{Proc. Am. Math. Sot.}, vol.~7, no.~1, 1956, pp.
  48--50.

\bibitem{797838}
P.-C. Chen, ``A non-line-of-sight error mitigation algorithm in location
  estimation,'' in \emph{Proc. Wireless Commun. Netw. Conf. (WCNC)}, Sep. 1999,
  pp. 316--3201.

\bibitem{4376962}
J.~Xing, J.~Zhang, L.~Jiao, X.~Zhang, and C.~Zhao, ``A robust wireless sensor
  network localization algorithm in {NLoS} environment,'' in \emph{Proc. IEEE
  Int. Conf. Control Automat. (ICCA)}, May 2007, pp. 3244--3249.

\bibitem{8481590}
C.~Qian, X.~Fu, N.~D. Sidiropoulos, and Y.~Yang, ``Tensor-based channel
  estimation for dual-polarized massive {MIMO} systems,'' \emph{IEEE Trans.
  Signal Process.}, vol.~66, no.~24, pp. 6390--6403, Dec. 2018.

\bibitem{golub1987generalization}
G.~H. Golub, A.~Hoffman, and G.~W. Stewart, ``A generalization of the
  eckart-young-mirsky matrix approximation theorem,'' \emph{Linear Algebra and
  its applications}, vol.~88, pp. 317--327, Apr. 1987, [Online]. Available:
  https://www.sciencedirect.com/science/article/pii/0024379587901145.

\bibitem{3GPP13810101}
3GPP, ``{5G}; {NR}; user equipment ({UE}) radio transmission and reception;
  part 1: Range 1 stand alone,'' {3rd Generation Partnership Project (3GPP)},
  Technical Specification (TS) 38.101, Jul. 2023, version 17.10.0.

\bibitem{ITU}
ITU, ``Calculation of free-space attenuation,'' {International Communication
  Union (ITU)}, Tech. Rep. P.524-4, Aug. 2019.

\bibitem{17564}
P.~Stoica and A.~Nehorai, ``{MUSIC}, maximum likelihood, and cramer-rao
  bound,'' \emph{IEEE Trans. Acoust. Speech Signal Process.}, vol.~37, no.~5,
  pp. 720--741, May 1989.

\bibitem{3GPP_DP}
3GPP, ``Study on channel model for frequencies from 0.5 to 100 {GHz},'' {3rd
  Generation Partnership Project (3GPP)}, Technical Specification (TS) 38.901,
  Dec. 2023, version 17.1.0.

\end{thebibliography}

\end{document}